\documentclass[graybox]{svmult}

\usepackage{mathptmx}       
\usepackage{makeidx}         
\usepackage{graphicx}        
\usepackage[bottom]{footmisc}
\usepackage{amsmath}       
\usepackage{url}
\usepackage{subfigure}

\begin{document}

\title{A Computation in a Cellular Automaton Collider Rule 110}
\author{Genaro J. Mart{\'i}nez, Andrew Adamatzky, and Harold V. McIntosh}
\institute{Unconventional Computing Centre, University of the West of England, \at Coldharbour Lane, Bristol BS16 1QY, UK 
\and Escuela Superior de C\'omputo, Instituto Polit\'ecnico Nacional, \at Av. Juan de Dios B\'atiz s/n, 07738, M\'exico 
\and Departamento de Aplicaci\'on de Microcomputadoras, Universidad Aut\'onoma de Puebla, \at 49 Poniente 1102, 72000, Puebla, Puebla, M\'exico} 

\maketitle

\abstract{A cellular automaton collider is a finite state machine build of rings of one-dimensional cellular automata. We show how a computation can be performed on the collider by exploiting interactions between gliders (particles, localisations). The constructions proposed are based on universality of elementary cellular automaton rule 110, cyclic tag systems, supercolliders, and computing on rings.}

\section{Introduction: Rule 110}
\label{rule110}

Elementary cellular automaton (CA) rule 110 \index{rule 110} is the binary cell state automaton with a local transition function $\varphi$ of a one-dimensional (1D) CA order $(k=2,r=1)$ in Wolfram's nomenclature \cite{cac}, where  $k$ is the number of cell states and $r$ the number of neighbours of a cell. We consider periodic boundaries, i.e. first and last cells of a 1D array are neighbours. The local transition function for rule 110 is defined in Tab.~\ref{rule110}, the string 01101110 is the number 110 in decimal notation:

\begin{align}
\varphi(1,1,1) \rightarrow 0 &\qquad{} \varphi(0,1,1) \rightarrow 1 \nonumber \\
\varphi(1,1,0) \rightarrow 1 &\qquad{} \varphi(0,1,0) \rightarrow 1 \nonumber \\
\varphi(1,0,1) \rightarrow 1 &\qquad {} \varphi(0,0,1) \rightarrow 1 \\
\varphi(1,0,0) \rightarrow 0 &\qquad {} \varphi(0,0,0) \rightarrow 0 \nonumber
\label{eqrule110}
\end{align}

\begin{figure}[th]
\centerline{\includegraphics[scale=.47]{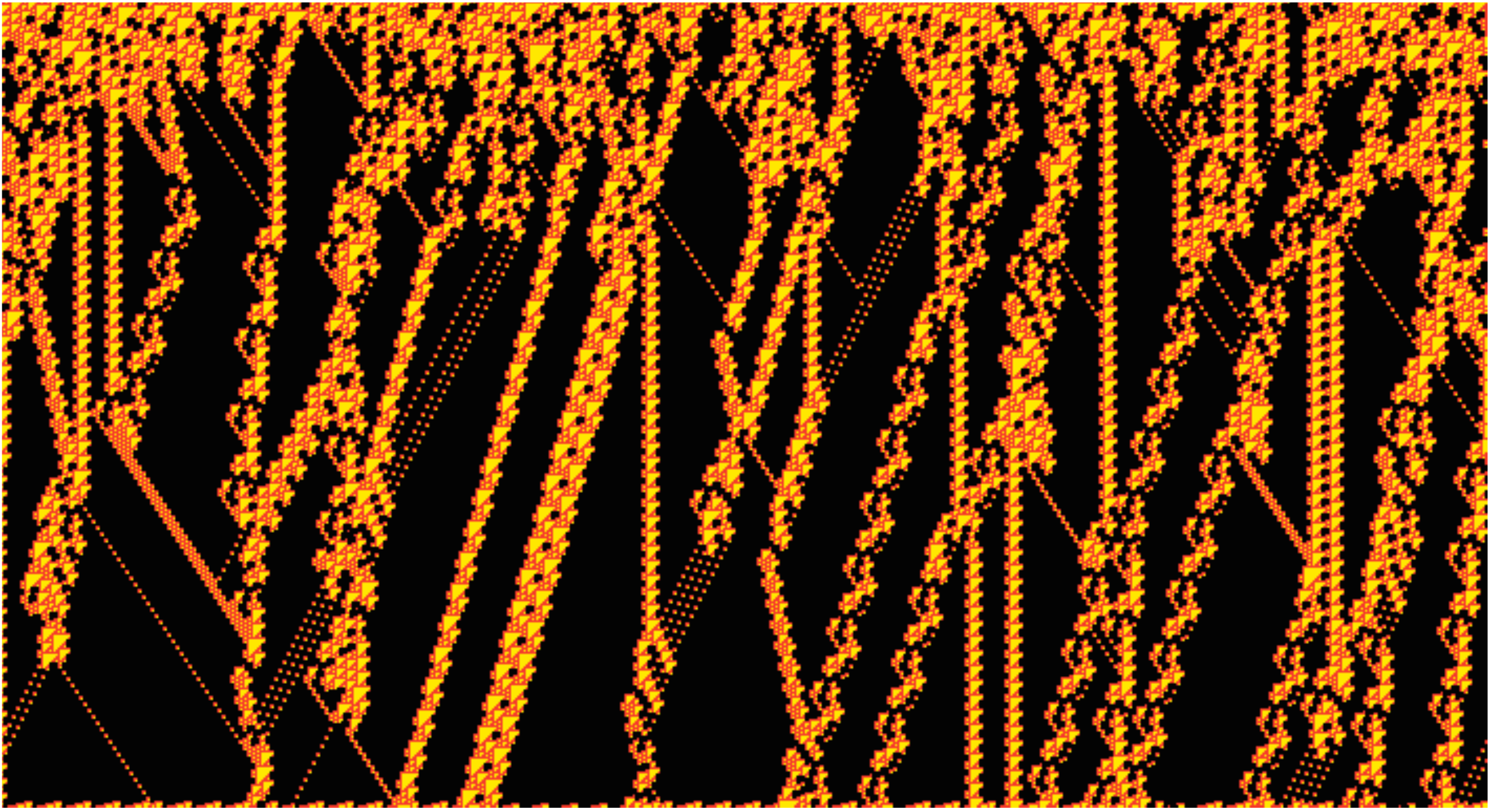}}
\caption{An example of CA rule 110 evolving for 384 time steps from a random configuration, where each cell assigned state `1' with uniformly distributed probability 0.5. The particles are filtered. Time goes down.}
\label{evolRandR110}
\end{figure}

A cell in state `0' takes state `1' if both its neighbours are in state `1' or left neighbour is `0' and right neighbour is `1'; otherwise, the call remains in the state `0'.  A cell in state `1' takes state `0' if both its neighbours are in state `1', or both its neighbours are in state `0' or it left neighbour is `1' and its right neighbour is `0'. Fig.~\ref{evolRandR110} shows an evolution of rule 110 from a random initial condition. We can see there travelling localisation: particles or gliders, and some stationary localisations: breathers, oscillators or stationary structures.

\subsection{System of particles}
\label{particlesrule110}

A detailed description of particles/gliders \index{particle} \index{glider} discovered in evolutions of CA rule 110 is provided in \cite{glidersmc, glidersIJUC}.\footnote{See also, \url{http://uncomp.uwe.ac.uk/genaro/rule110/glidersRule110.html}} Further, we refers to a train of $n$ copies of particle $A$ as $A^n$. 

\begin{figure}[th]
\centerline{\includegraphics[scale=.46]{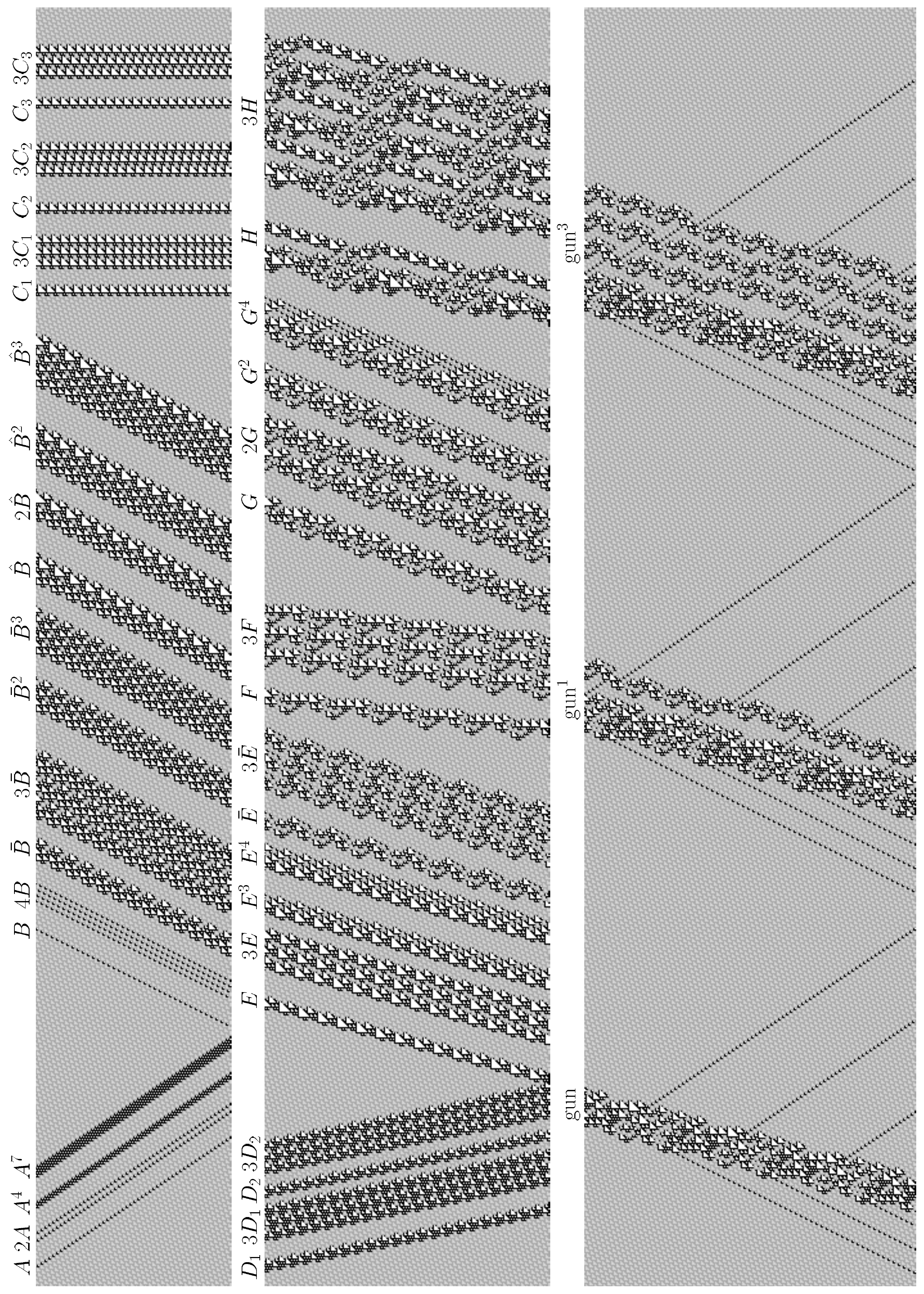}}
\caption{Types of particles discovered in rule 110.}
\label{listaCookgliders}
\end{figure}

\begin{table}[th]
\caption{Properties of particles in rule 110.}
\label{margenesgliders}
\begin{tabular}{ccccccc}
\hline\noalign{\smallskip}
                 & \multicolumn{4}{c}{Margins} & & \\
Structure & \multicolumn{4}{c}{Left -- Right} & Velocity & Lineal \\
\cline{2-5}
 & $ems$ & $oms$ & $ems$ & $oms$ & & Volume \\
\noalign{\smallskip}\svhline\noalign{\smallskip}
$e_{r}$ & . & 1 & . & 1 & 2/3 $\approx$ 0.666666 & 14  \\
$e_{l}$ & 1 & . & 1 & . & -1/2 = -0.5 & 14 \\
$A$ & . & 1 & . & 1 & 2/3 $\approx$ 0.666666 & 6  \\
$B$ & 1 & . & 1 & . & -2/4 = -0.5 & 8 \\
$\bar{B}^{n}$ & 3 & . & 3 & . & -6/12 = -0.5 & 22 \\
$\hat{B}^{n}$ & 3 & . & 3 & . & -6/12 = -0.5 & 39 \\
$C_{1}$ & 1 & 1 & 1 & 1 & 0/7 = 0 & 9-23 \\
$C_{2}$ & 1 & 1 & 1 & 1 & 0/7 = 0 & 17 \\
$C_{3}$ & 1 & 1 & 1 & 1 & 0/7 = 0 & 11 \\
$D_{1}$ & 1 & 2 & 1 & 2 & 2/10 = 0.2 & 11-25 \\
$D_{2}$ & 1 & 2 & 1 & 2 & 2/10 = 0.2 & 19 \\
$E^{n}$ & 3 & 1 & 3 & 1 & -4/15 $\approx$ -0.266666 & 19 \\
$\bar{E}$ & 6 & 2 & 6 & 2 & -8/30 $\approx$ -0.266666 & 21 \\
$F$ & 6 & 4 & 6 & 4 & -4/36  $\approx$ -0.111111 & 15-29 \\
$G^{n}$ & 9 & 2 & 9 & 2 & -14/42 $\approx$ -0.333333 & 24-38 \\
$H$ & 17 & 8 & 17 & 8 & -18/92 $\approx$ -0.195652 & 39-53 \\
glider gun & 15 & 5 & 15 & 5 & -20/77 $\approx$ -0.259740 & 27-55 \\
\noalign{\smallskip}\hline\noalign{\smallskip}
\end{tabular}
\end{table}

Figure~\ref{listaCookgliders} shows all known particles, and generators of particles, or glider guns.\index{glider!gun} Each particle has its unique features, e.g. slopes, velocities, periods, contact points, collisions, and phases \cite{reglangr110, atlasr110, concordmc}. A set of particles in rule 110 is defined as:

$$
{\texttt G}=\{A, B, \bar{B}^n, \hat{B}^{n}, C_1, C_2, C_3, D_1, D_2, E^n, \bar{E}, F, G^n, H, gun^n\}.
$$

\noindent where $n$ means that a structure of the particle can be extendible infinitely, the rest of symbols denote types of particles as shown in Fig.~\ref{listaCookgliders}. Table~\ref{margenesgliders} summarizes  key features of the particles: column {\it structure} gives the name of each particle including two more structures: $e_{r}$ and $e_{l}$ which represent the slopes of ether pattern (periodic background). The next four columns labeled {\it margins} indicate the number of periodic margins in each particle: they are useful to recognize contact points for collisions. The margins are partitioned in two types with even values {\it ems} and odd values {\it oms} which are distributed also in two groups: left and right margins. Column $v_g$ indicates a velocity of a particle $g$, where $g$ belongs to a particle of the set of particles $\texttt G$. A relative velocity is calculated during the particle's displacement on $d$ cells during period $p$. We indicate three types of a particle propagation via sign of its velocity. A particle travelling to the right has {\it positive velocity}, a particle travelling to the left has {\it negative velocity}. Stationary particle has zero velocity. Different velocities of particles allow us to control distances between the particle to obtained programmable reactions between the particles. Typically, larger particles has lower velocity values. No particle can move faster than $v_{e_r}$ or $v_{e_l}$. Column {\it lineal volume} shows the minimum and maximum number of necessary cells occupied by the particle.

\subsection{Particles as regular expressions}

We represent CA particles as strings. These strings can be calculated using de Bruin diagrams \cite{deBruijnmc, mcbook, cavoorhees, glidersmc, reglangr110} \index{de Bruin diagram} or with the tiles theory \cite{tilesbook, concordmc, atlasr110, reglangr110}. \footnote{See a complete set of regular expressions for every particle in rule 110 in \url{http://uncomp.uwe.ac.uk/genaro/rule110/listPhasesR110.txt}}

A regular language $L_{R110}$ is based on a set of regular expressions $\Psi_{R110}$ uniquely describing every particle of $\texttt G$. A subset of the set of regular expressions 

\begin{equation}
\Psi_{R110} = \bigcup_{i=1}^{p}{w_{i,g}} \mbox{ } \forall \mbox{ } (w_i \in \Sigma^* \wedge g \in \texttt G)
\end{equation}

\noindent where  $p \geq 3$ is a period,  determines the language

\begin{equation}
L_{R110} = \{w | w = w_iw_j \vee w_i+w_j \vee w_i^* \mbox{ and } w_i, w_j \in \Psi_{R110}\}.
\end{equation}

From these set of strings we can code initial configurations to program collisions between particles~\cite{objr110, ecasolitons, glidersIJUC}.

\begin{figure}[th]
\centerline{\includegraphics[scale=.535]{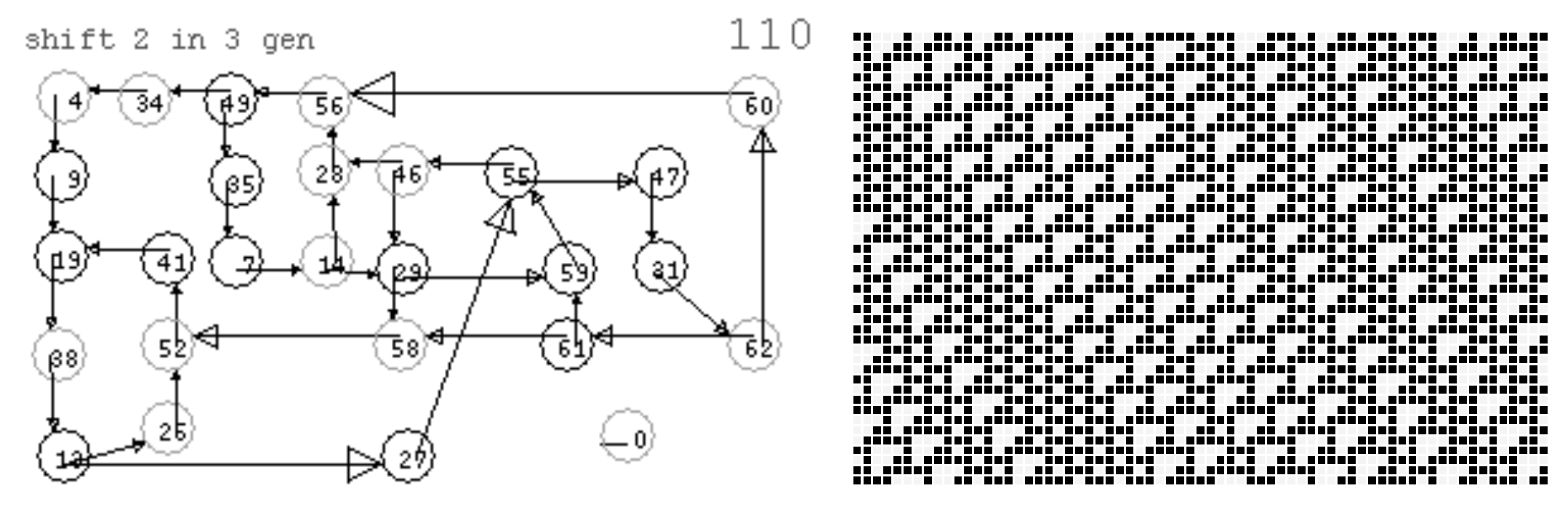}}
\caption{De Bruijn diagram calculating $A$ particles (left) and space-time configuration of automaton showing locations of periodic sequences produced (right).}
\label{fases-gliderA}
\end{figure}

To deriver the regular expressions \index{regular expression} we use the de Bruijn diagrams \cite{deBruijnmc, cavoorhees, mcbook} as follows. Assume the particle $A$ moves two cells to the right in three time steps (see Tab.~\ref{margenesgliders}). The corresponding extended de Bruijn diagram (2-shift, 3-gen) is shown in Fig.~\ref{fases-gliderA}. Cycles in the diagram are periodic sequences uniquely representing each phase of the particle. Diagram in Fig.~\ref{fases-gliderA} has two cycles: a cycle formed by just a vertex 0 and another large cycle of 26 vertices composed by other nine internal cycles. The sequences or regular expressions determining the phases of the particle $A$ are obtained by following paths through the edges of the diagram. There regular expressions and corresponding paths in Bruijn diagram are shown below.

\begin{quotation}
\begin{enumerate}
\item[I.] The expression (1110)*: vertices 29, 59, 55, 46 determining $A^{n}$ particles.
\item[II.] The expression (111110)*: vertices 61, 59, 55, 47, 31, 62  defining $nA$ particles with a $T_{3}$ tile among each particle.
\item[III.] The expression (11111000100110)*: vertices 13, 27, 55, 47, 31, 62, 60, 56, 49, 34, 4, 9, 19, 38 describing the periodic background configurations in a specific phase.
\end{enumerate}
\end{quotation}

Cycle with period 1 (vertex 0) yields a homogeneous evolution in state 0. The evolution space in Fig.~\ref{fases-gliderA} shows different trains of $A$ particles. The initial condition is constructed following some of the seven possible cycles of the de Bruijn diagram or a combination of them. In this way, the number of particles $A$ or the number of intermediate tiles $T_{3}$ can be selected by moving from one cycle to another.

The alignment of the f$_{i}$\_1 phases is analysed to determine the whole set of strings for every particle. We describe the form and limits of each particle by tiles. Then a phase is fixed (in our case the phase f$_{i}$\_1) and a horizontal line is placed in the evolution space bounded by two aligned $T_{3}$ tiles. The sequence between both tiles aligned in each of the four levels determines a periodic sequence representing a particular structure in the evolution space of rule 110. All periodic sequences in a specific phase are calculated, enumerating the phases for each particle or non-periodic structure.

\begin{table}[th]
\caption{Four sets of phases $Ph_i$ in rule 110.}
\begin{tabular}{rcc}
phases level one ($Ph_1$) & $\rightarrow$ & $\{$f$_{1}$\_1, f$_{2}$\_1, f$_{3}$\_1, f$_{4}$\_1$\}$ \\
phases level two ($Ph_2$) & $\rightarrow$ & $\{$f$_{1}$\_2, f$_{2}$\_2, f$_{3}$\_2, f$_{4}$\_2$\}$ \\
phases level three ($Ph_3$) & $\rightarrow$ & $\{$f$_{1}$\_3, f$_{2}$\_3, f$_{3}$\_3, f$_{4}$\_3$\}$ \\
phases level four ($Ph_4$) & $\rightarrow$ & $\{$f$_{1}$\_4, f$_{2}$\_4, f$_{3}$\_4, f$_{4}$\_4$\}$
\end{tabular}
\label{fases-fi_i}
\end{table}

Table~\ref{fases-fi_i} represents disjoint subset of phases, each level contains four phases. Variable f$_{i}$ indicates the phase of a particle, and the subscript $j$ (in the notation f$_{i}$\_$j$) indicates the selected set $Ph_j$ of regular expressions. Finally, we use the next notation to codify initial conditions by phases as follows:

\begin{equation}
\#_{1}(\#_{2},\mbox{f}_{i}\_1)
\end{equation}

\noindent where \#$_{1}$ represents a particle according to Cook's classification (Table~\ref{margenesgliders}) and \#$_{2}$ is a phase of the particle with period greater than four.

\section{Universal elementary CA }

A concept of universality \index{universality} and self-reproduction \index{self-reproduction} in CA  was proposed by von Neumann in \cite{von} in his design of a universal constructor in a 2D CA with 29 cell-states. Architectures of universal CA have been simplified by Codd in 1968 \cite{coddbook}, Banks in 1971 \cite{banksthesis}, Smith in 1971 \cite{smithca}, Conway in 1982 \cite{ugol}, Lindgren and Nordahl in 1990 \cite{uoca}, and Cook in 1998 \cite{ecauniversality}.\footnote{A range of universal CA is listed here \url{http://uncomp.uwe.ac.uk/genaro/Complex_CA_repository.html}} Cook simulated a cyclic tag system, equivalent to a minimal Turing machine, in CA rule 110. In general, computation capacities are explores in complex CA and chaotic CA \cite{compcaclass}.

\section{Cyclic tag systems}

Cyclic tag systems \index{cyclic tag system} are used by Cook in \cite{ecauniversality} as a tool to implement computations in rule 110. Cyclic tag systems  are modified from tag systems \index{tag system} by allowing the system to have the same action of reading a tape in the front and adding characters at its end:

\begin{quotation}
\begin{enumerate}
\item Cyclic tag systems have at least two letters in their alphabet ($\mu > 1$).
\item Only the first character is deleted ($\nu = 1$) and its respective sequence is added.
\item In all cases if the machine reads a character zero then the production rule is always null ($0 \rightarrow \epsilon$, where $\epsilon$ represents the empty word).
\item There are $k$ sequences from $\mu^*$ which are periodically accessed to specify the current production rule when a nonzero character is taken by the system. Therefore the period of each cycle is determinate by $k$.
\end{enumerate}
\end{quotation}

Such cycle determines a partial computation over the tape,  although a halt condition is not specified. Let us see some samples of a cyclic tag system working with $\mu=2$, $k=3$ and the following production rules: $1 \rightarrow 11$, $1 \rightarrow 10$ and $1 \rightarrow \epsilon$. To avoid writing a chain when there is no need to add characters, the $\vdash_k$ relation is just indicated. For example, the 00001 $\vdash_1 \vdash_2 \vdash_3 \vdash_1 \vdash_2 10$ represents the relations 00001 $\vdash_1$ 0001 $\vdash_2$ 001 $\vdash_3$ 01 $\vdash_1$ 1 $\vdash_2$ 10. Each relation indicates which exactly sequence $\mu$ is selected.

Cyclic tag systems tend to growth quickly which makes it difficult to analyse their behaviour. Morita in \cite{urca, tmcts} demonstrated how to implement a particular halt condition in cyclic tag systems given an output string when the system is halting, and how a partitioned CA can simulate any cyclic tag system, consequently computing all the recursive functions.

Similar to Post's developments with tag systems, Cook determined that for a cyclic tag system with $\mu=2$, $k=2$, the productions $1 \rightarrow 11$ and $1\rightarrow 10$, and starting evolution with the state 1 on the tape, it is impossible to decide if the process is terminal.

\section{Cyclic tag system working in rule 110}

Let us see how a cyclic tag system operates in rule 110 \cite{ankos}. We use a cyclic tag system with $\mu=2$, $k=2$ and the productions $1 \rightarrow 11$ and $1\rightarrow 10$, starting its evolution in state 1 on the tape. A fragments of the systems' behaviour is shown below:

\begin{quotation}
1 $\vdash_1$ 11 $\vdash_2$ 110 $\vdash_1$ 1011 $\vdash_2$ 01110 $\vdash_1 \vdash_2$ 11010 $\vdash_1$ 101011 $\vdash_2$ 0101110 $\vdash_1 \vdash_2$ 0111010 $\vdash_1 \vdash_2$ 1101010 $\vdash_1$ 10101011 $\vdash_2$ 010101110 $\vdash_1 \vdash_2$ 010111010 $\vdash_1 \vdash_2$ 011101010 $\vdash_1 \vdash_2$ 110101010 $\vdash_1$ 1010101011 $\vdash_2$ 01010101110 $\vdash_1 \vdash_2$ 01010111010 $\vdash_1 \vdash_2$ 01011101010 $\vdash_1 \vdash_2$ 01110101010 $\vdash_1 \vdash_2$ 11010101010 $\vdash_1$ 101010101011 $\vdash_2$ 0101010101110 $\vdash_1 \vdash_2$ 0101010111010 $\vdash_1 \vdash_2$ 01010111010
10 $\vdash_1 \vdash_2$ 0101110101010 $\vdash_1 \vdash_2$ 0111010101010 $\vdash_1 \vdash_2$ 1101010101010 $\vdash_1$ 10101010101011 $\vdash_2$ 010101010101110 $\vdash_1 \vdash_2$ 010101010111010 $\vdash_1 \vdash_2$ 01010
1011101010 $\vdash_1 \vdash_2$ 010101110101010 $\vdash_1 \vdash_2$ 010111010101010 $\ldots$
\end{quotation}

We start with the expression 1(10)*. The cyclic tag systems moves (from the right to the left) and adds a pair of bits. As soon as the expression 1(10)* appears again, a number of relations selected in each interval in such a manner that the expressions grow lineally in order of $f_1=2(n+1)$.

\begin{figure}[th]
\centerline{\includegraphics[scale=.58]{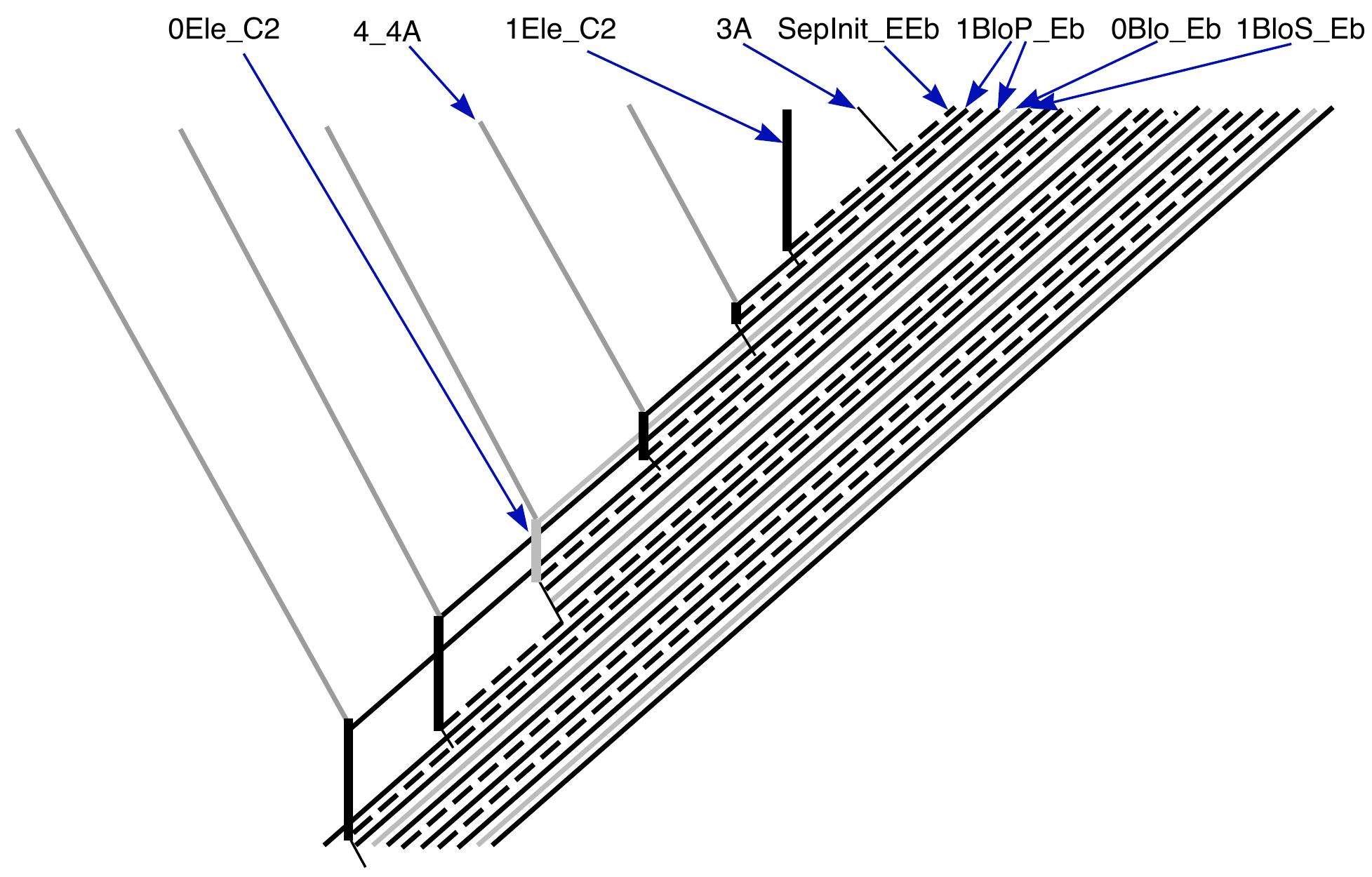}}
\caption{Schematic diagram of a cyclic tag system working in rule 110.}
\label{CTS-diagrama}
\end{figure}

If we take consecutive copies of 1(10)* with their respective intervals determined by the number of $j$ productions (represented as $\vdash_i^{j}$), we obtain the following sequence: 1 $\vdash_i^{2}$ 110 $\vdash_i^{4}$ 11010 $\vdash_i^{6}$ 1101010 $\vdash_i^{8}$ 1101010 $\vdash_i^{10}$ 110101010 $\vdash_i^{12}$ 11010101010 $\vdash_i^{14}$ 1101010101010 $\vdash_i^{16} \ldots$. There are no states where to `0' appear together. 

Further, we show how to interpret particles and their collisions to emulate a cyclic tag system in rule 110. We must use trains of particles to represent data and operators, their reactions, transform and deletion of data on the tape. A schematic diagram, where trains of particles are represented by lines, is shown in Fig.~\ref{CTS-diagrama}. The diagram is explained with details in the next sections.

\subsection{Components based on sets of particles}
\label{compcts}

A construction of the cyclic tag system in rule 110 can be subdivided into three parts (Fig.~\ref{CTS-diagrama}). First part is the left periodic part controlled by trains of 4\_$A^{4}$ particles. This part is static. It controls the production of $0$'s and $1$'s. The second part is the center determining the initial value on the tape. The third part is the right, cyclic, part which contains the data to process. It adds or removes data on the tape.

\newpage

\begin{center}
{\bf Set of particles 4\_$A^{4}$}
\end{center}

The four trains of $A^{4}$ particles are static but their phases change periodically. A key point is to implement these components by defining both distances and phases, because some choices of phases or distances might induce an undesirable reactions between the trains of particles.

\begin{figure}[th]
\centerline{\includegraphics[scale=.86]{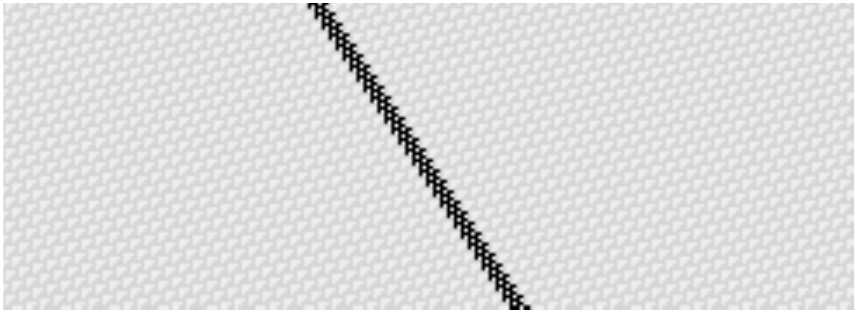}}
\caption{Set of particles 4\_$A^{4}$.}
\label{4As}
\end{figure}

Packages defined by particles $A^{4}$ have three different phases: f$_{1}$\_1, f$_{2}$\_1 and f$_{3}$\_1. To construct the first train 4\_$A^{4}$ we must establish the phase of each $A^{4}$. Let us assign phases as follows: \\

\noindent $A^{4}$(f$_{3}$\_1)-$27e$-$A^{4}$(f$_{2}$\_1)-$23e$-$A^{4}$(f$_{1}$\_1)-$25e$-$A^{4}$(f$_{3}$\_1), \\

\noindent see Fig.~\ref{4As}. Spaces between each train 4\_$A^{4}$ are fixed but the phases change. The soliton-like collisions between the particles $\bar{E}$ occur: \\

\noindent \{$649e$-$A^{4}$(f$_{2}$\_1)-$27e$-$A^{4}$(f$_{1}$\_1)-$23e$-$A^{4}$(f$_{3}$\_1)-$25e$-$A^{4}$(f$_{2}$\_1)-$649e$-$A^{4}$(f$_{1}$\_1)-\\
$27e$-$A^{4}$(f$_{3}$\_1)-$23e$-$A^{4}$(f$_{2}$\_1)-$25e$-$A^{4}$(f$_{1}$\_1)]-$649e$-$A^{4}$(f$_{3}$\_1)-$27e$-$A^{4}$(f$_{2}$\_1)-$23e$-\\
$A^{4}$(f$_{1}$\_1)-$25e$-$A^{4}$(f$_{3}$\_1)\}* \\

\noindent If for every 4\_$A^{4}$ we take a phase representing the complete train, we can rename it as:

\begin{center}
\{$649e$-4\_$A^{4}$(F$_{2}$)-$649e$-4\_$A^{4}$(F$_{1}$)-$649e$-4\_$A^{4}$(F$_{3}$)\}*
\end{center}

\noindent this phase change is important to preserve good reactions coming to the left side of the system. \\ \\

\begin{center}
{\bf Set of particles 1Ele$\_C_{2}$ and 0Ele$\_C_{2}$}
\end{center}

The central part is made of the state `1' written on the tape represented by a train of four $C_{2}$ particles. A set of particles 1Ele$\_C_{2}$ represents `1' and a set of particles 0Ele$\_C_{2}$ represents `0' on the tape.

\begin{figure}[th]
\centerline{\includegraphics[scale=.85]{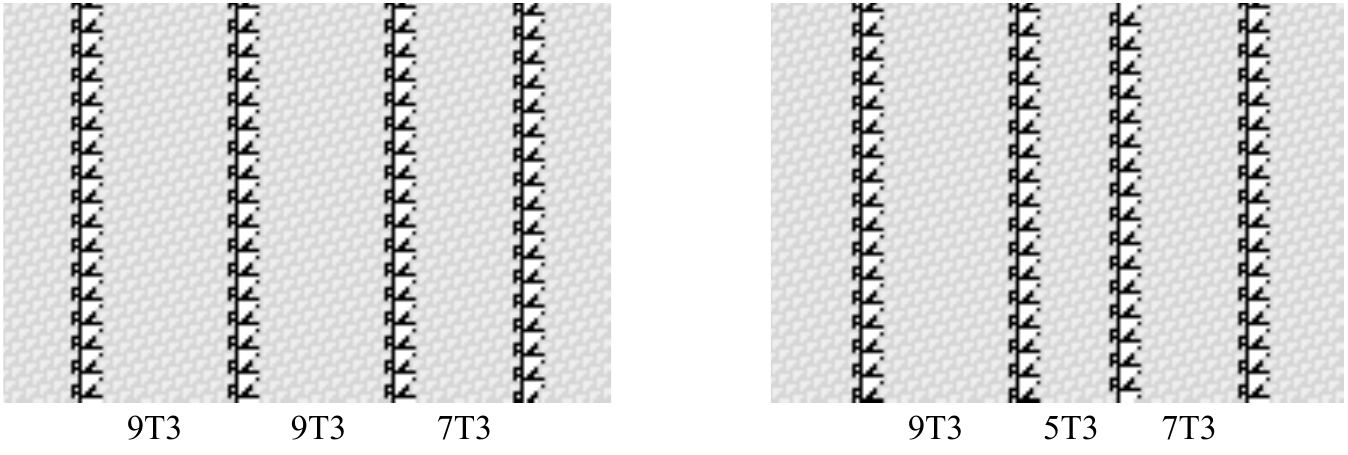}}
\caption{Set of particles 1Ele$\_C_{2}$ (left) and 0Ele$\_C_{2}$ (right).}
\label{1-0-elementos}
\end{figure}

The left configurations in Fig.~\ref{1-0-elementos} shows the set of particles 1Ele$\_C_{2}$. We should reproduce each set of particles by the phases f$_{i}$\_1. The phases are coded as follows: $C_{2}$(A,f$_{1}$\_1)-$2e$-$C_{2}$(A,f$_{1}$\_1)-$2e$-$C_{2}$(A,f$_{1}$\_1)-$e$-$C_{2}$(B,f$_{2}$\_1). The first three particles $C_{2}$ are in phase (A,f$_{1}$\_1) and the fourth particle $C_{2}$ is in phase (B,f$_{2}$\_1). The distances between the particles are  $9T_{3}$-$9T_{3}$-$7T_{3}$. To determine the distances,  we count the number of tiles $T_{3}$ between particles. Similarly, we obtain the distances   $9T_{3}$-$5T_{3}$-$7T_{3}$ for the particles 0Ele$\_C_{2}$.

\begin{center}
{\bf Set of particles 0Blo$\_\bar{E}$}
\end{center}

The left part stores blocks of data without transformations in trains of $E$ and the particles $\bar{E}$.

\begin{figure}[th]
\centerline{\includegraphics[scale=.65]{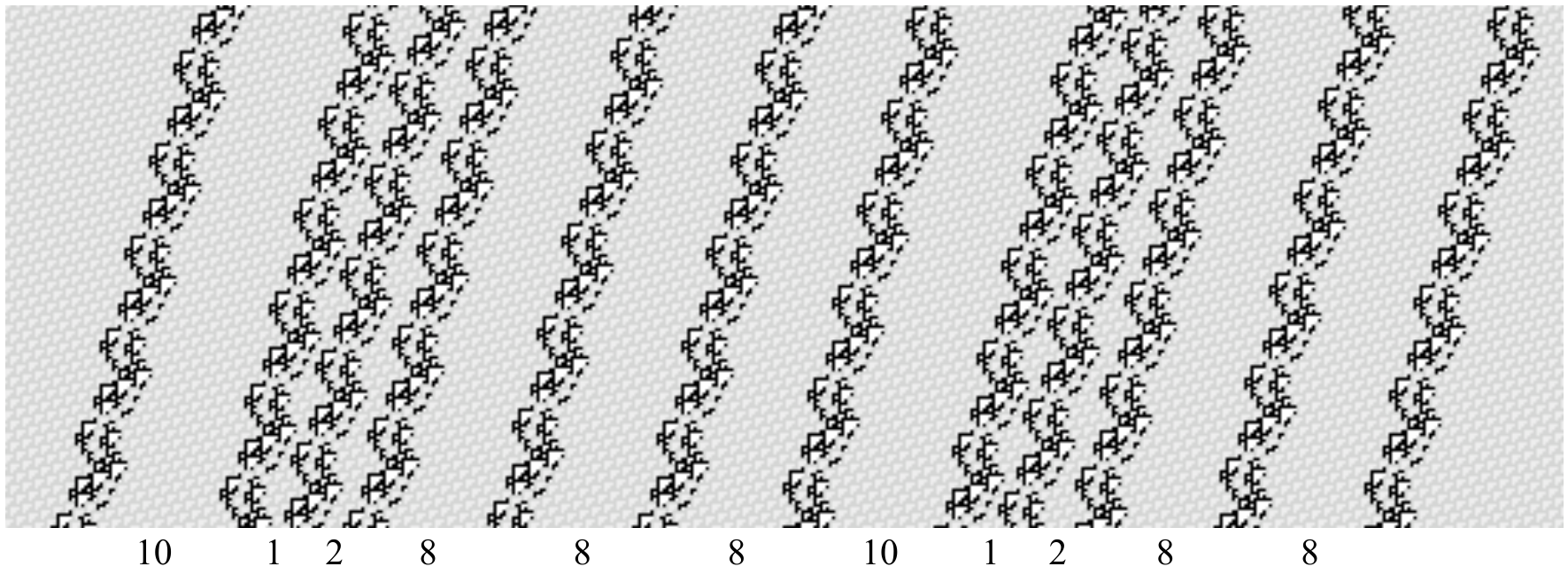}}
\caption{Set of particles 0Blo$\_\bar{E}$.}
\label{0Blo_Eb-elemento}
\end{figure}

The set of particles 0Blo$\_\bar{E}$ is formed by 12$\bar{E}$ particles as we can see in Fig.~\ref{0Blo_Eb-elemento}. There must be an exact phase and distance between each one of the particles, otherwise the whole system will be disturbed. 

\begin{center}
{\bf Set of particles 1BloP$\_\bar{E}$ and 1BloS$\_\bar{E}$}
\end{center}

To write `1's we must use two set of particles --- {\it primary} and {\it standard}.

\begin{figure}[th]
\centerline{\includegraphics[scale=.65]{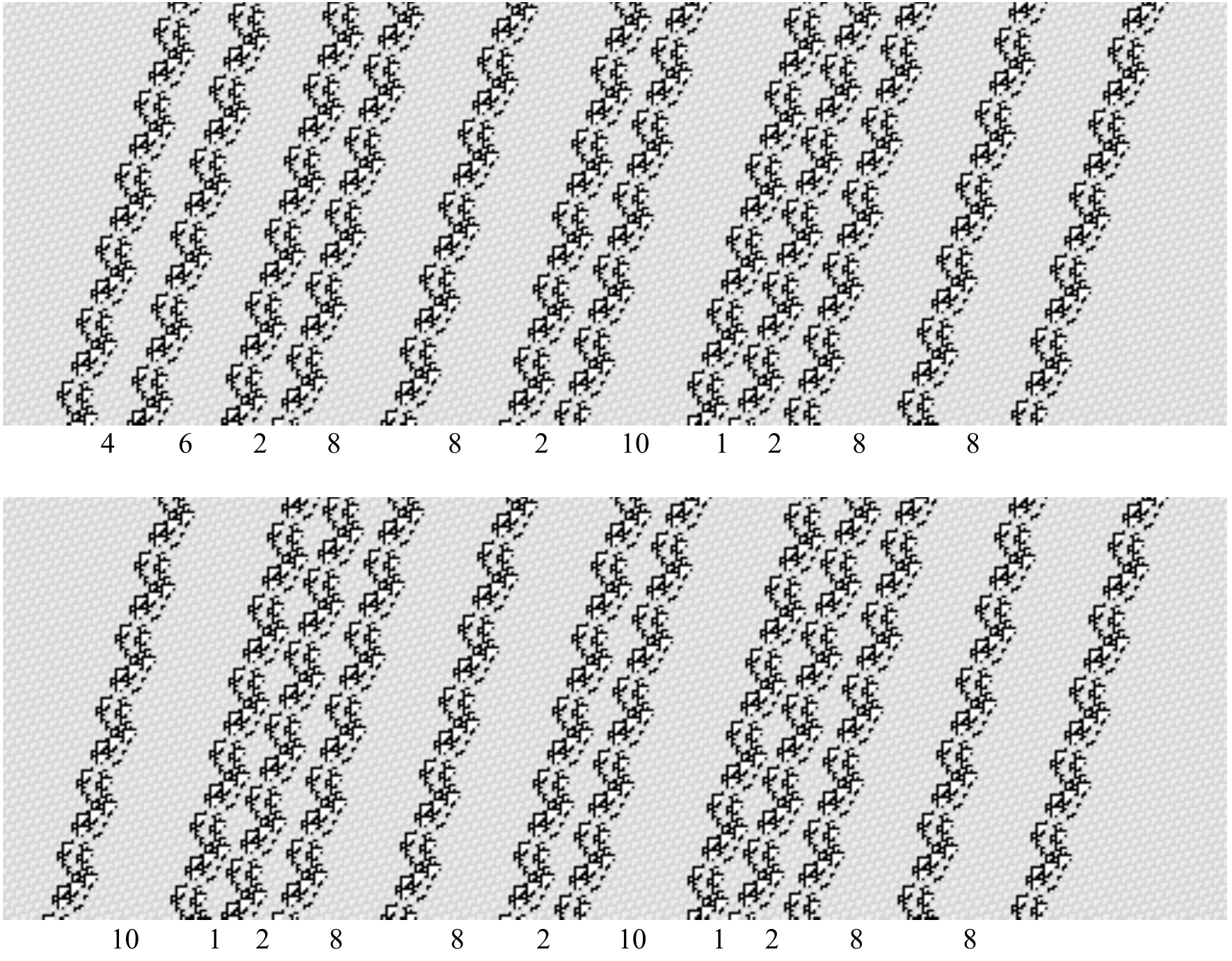}}
\caption{Set of particles 1BloP$\_\bar{E}$ (left) and 1BloS$\_\bar{E}$ (right).}
\label{1Blo_Eb-elementos}
\end{figure}

They are differences in distance between first two particles $\bar{E}$, as shown in Fig.~\ref{1Blo_Eb-elementos}. Both blocks produce the same set of particles 1Add$\_\bar{E}$. The main reason to use both set of particles is because the CA rule 110 evolves asymmetrically and therefore we need a double set of particles to produce values 1 correctly.

\begin{center}
{\bf Set of particles SepInit$\_E\bar{E}$}
\end{center}

\begin{figure}[th]
\centerline{\includegraphics[scale=.72]{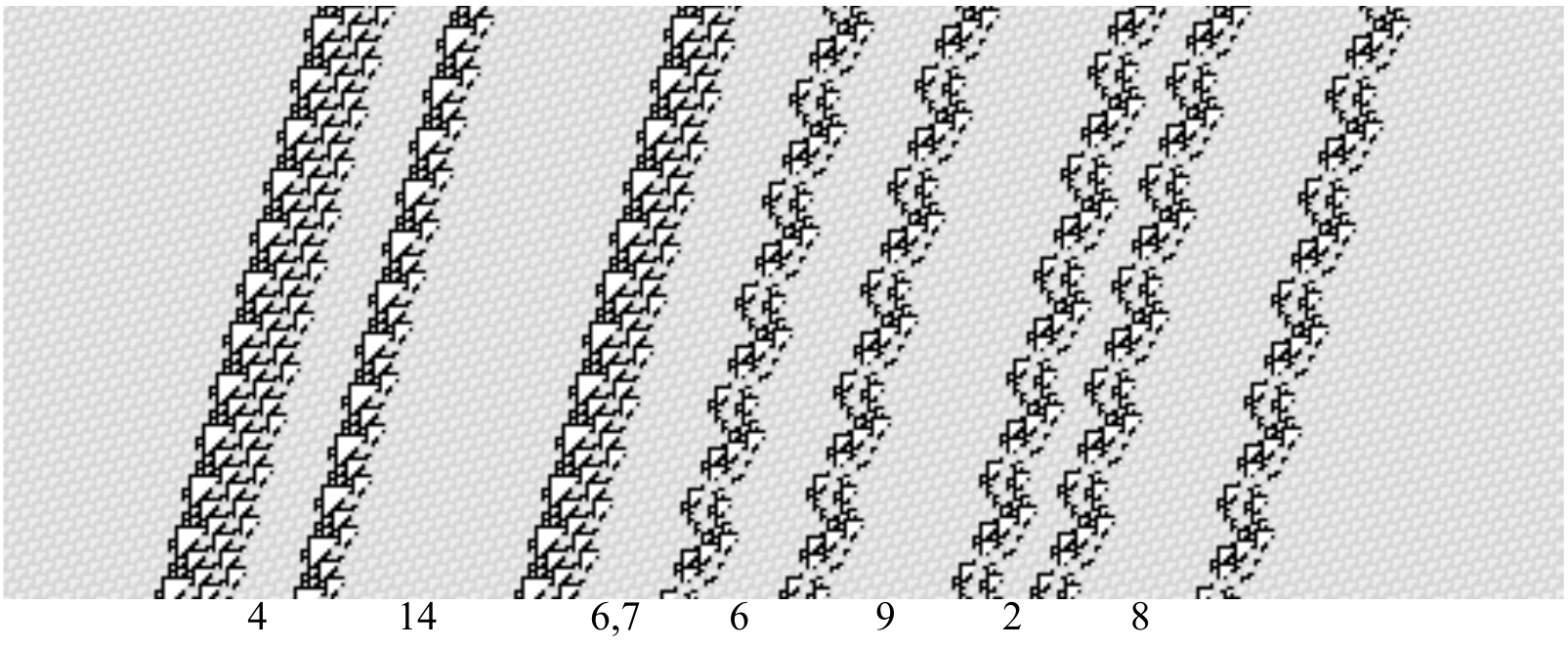}}
\caption{Set of particles SepInit$\_E\bar{E}$.}
\label{SepInit_EEb-elemento}
\end{figure}

A leader component renamed as the set of particles SepInit$\_E\bar{E}$ (see Fig.~\ref{SepInit_EEb-elemento}) is essential to separate trains of data and to determine the incorporation of the data on the tape. Its has a small but detailed code determining which data without transformation would be added or erased from the tape, depending on the value that is coming.

\begin{center}
{\bf Set of particles 1Add$\_\bar{E}$ and 0Add$\_\bar{E}$}
\end{center}

\begin{figure}[th]
\centerline{\includegraphics[scale=.65]{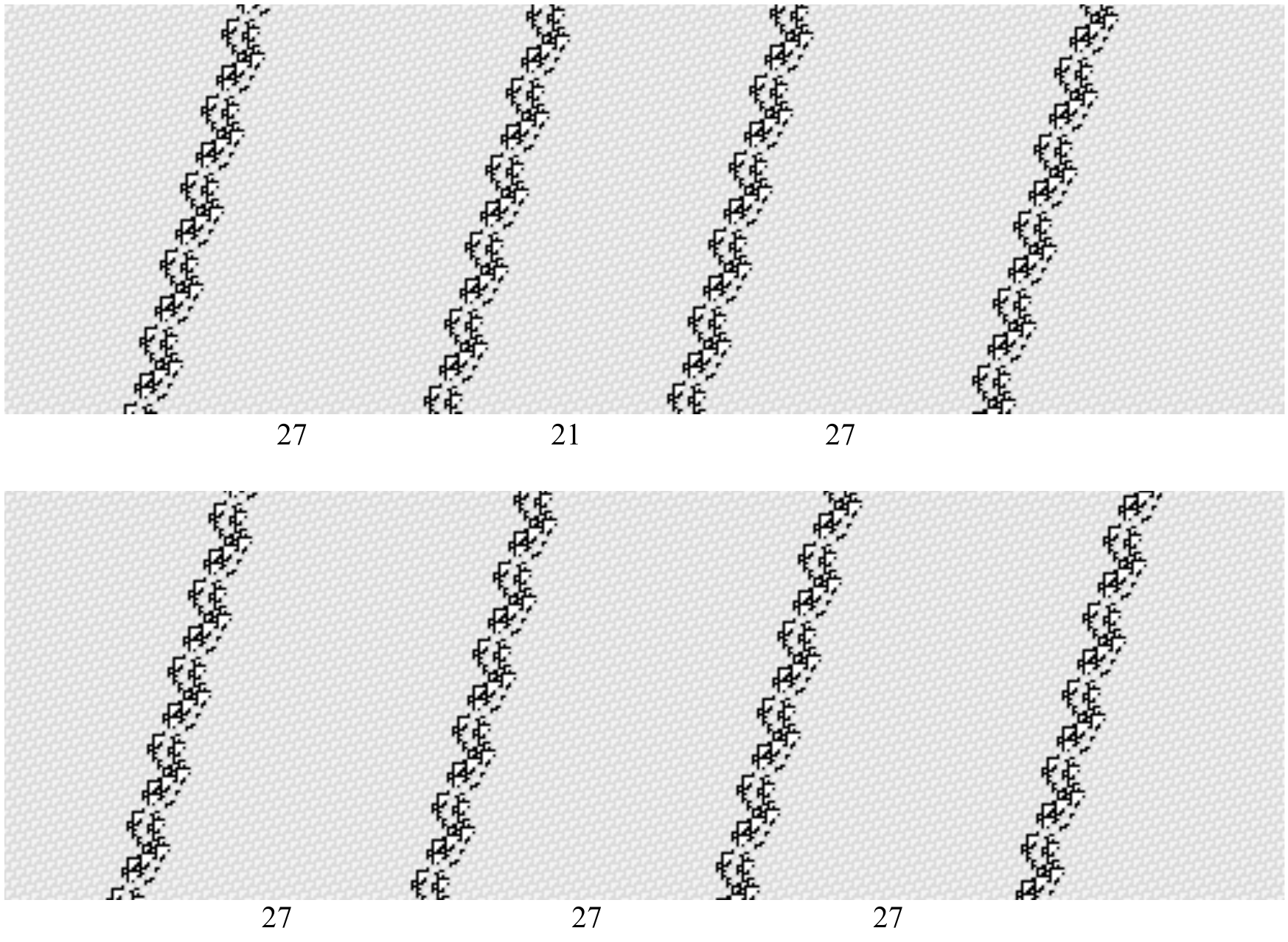}}
\caption{Set of particles 1Add$\_$Eb (left) and 0Add$\_\bar{E}$ (right).}
\label{Add_Eb-elementos}
\end{figure}

Figure~\ref{Add_Eb-elementos} illustrates the set of particles 1Add$\_\bar{E}$ and 0Add$\_\bar{E}$ produced by two previous different trains of data. A set of particles 1Add$\_\bar{E}$ must be generated by the set of particles 1BloP$\_\bar{E}$ or 1BloS$\_\bar{E}$. This way, both set of particles can produce the same element.

On the other hand, a set of particles 0Add$\_\bar{E}$ is generated by a set of particles 0Blo$\_\bar{E}$. Nevertheless, we could produce $\bar{E}$ particles modifying their first two distances and preserving them without changing others particles to get a reliable reaction. This is possible if we want to experiment with other combinations of blocks of data.

If a leader set of particles SepInit$\_E\bar{E}$ reaches a set of particles 1Ele$\_\bar{E}$, it erases this value from the tape and adds a new data that shall be transformed. In other case, if it finds a set of particles 0Ele$\_\bar{E}$, then it erases this set of particles from the tape and also erases a set of unchanged data which comes from the right until finding a new leader set of particles. This operation represents the addition of new values from periodic trains of particles coming from the right. Thus a set of particles 1Add$\_\bar{E}$ is transformed into 1Ele$\_\bar{E}$ colliding against a train of 4$\_A^{4}$ particles representing a value 1 in the tape, and the set of particles 0Add$\_\bar{E}$ is transformed into 0Ele$\_\bar{E}$ colliding against a train of 4$\_A^{4}$ particles representing a value 0 in the tape.

\begin{table}
\caption{Distances between sets of particles.}
\begin{tabular}{l|l}
set of particles & distance \\
\hline
1Ele$\_C_{2}$ & 9-9-7 \\
0Ele$\_C_{2}$ & 9-5-7 \\
1BloP$\_\bar{E}$ & 4-6-2-8-8-2-10-1-2-8-8 \\
1BloS$\_\bar{E}$ & 10-1-2-8-8-2-10-1-2-8-8 \\
0Blo$\_\bar{E}$ & 10-1-2-8-8-8-10-1-2-8-8 \\
SepInit$\_E\bar{E}$ & 4-14-$(6 \mbox{ or } 7)$-6-9-2-8 \\
1Add$\_\bar{E}$ & 27-21-27 \\
0Add$\_\bar{E}$ & 27-27-27
\end{tabular}
\label{distancias}
\end{table}

Table~\ref{distancias} shows all distances (in numbers of $T_3$ tiles) for every. We can code the construction of this cyclic tag system across phase representations in three main big sub systems:

\begin{quotation}
\begin{tabbing}
{\bf left:} \ldots \= -$217e$-4$\_A^{4}$(F2)-$649e$-4$\_A^{4}$(F1)-$649e$-4$\_A^{4}$(F3)-$649e$-4$\_A^{4}$(F2)- \\ 
\> $649e$-4$\_A^{4}$(F1)-$649e$-4$\_A^{4}$(F3)-$216e$- \\ \\
{\bf center:} 1Ele$\_C_{2}$(A,f$_{1}$\_1)-$e$-$A^{3}$(f$_{1}$\_1)- \\ \\
{\bf right:} \= SepInit$\_E\bar{E}$(C,f$_{3}$\_1)-1BloP$\_\bar{E}$(C,f$_{4}$\_1)-SepInit$\_E\bar{E}$(C,f$_{3}$\_1)- \\
\> 1BloP$\_\bar{E}$(C,f$_{4}$\_1)-0Blo$\_\bar{E}$(C,f$_{4}$\_1)-1BloS$\_\bar{E}$(A,f$_{4}$\_1)- \\
\> SepInit$\_E\bar{E}$(A,f$_{2}$\_1)(2)-1BloP$\_\bar{E}$(F,f$_{1}$\_1)-SepInit$\_E\bar{E}$(A,f$_{3}$\_1)(2)- \\
\> 1BloP$\_\bar{E}$(F,f$_{1}$\_1)-0Blo$\_\bar{E}$(E,f$_{4}$\_1)-1BloS$\_\bar{E}$(C,f$_{4}$\_1)-$e$- \\
\> SepInit$\_E\bar{E}$(B,f$_{1}$\_1)(2)-1BloP$\_\bar{E}$(F,f$_{3}$\_1)-$e$- \\
\> SepInit$\_E\bar{E}$(B,f$_{1}$\_1)(2)-$217e$-\dots.
\end{tabbing}
\end{quotation}

The initial conditions in rule 110 are able to generate the serial sequence of bits 1110111 and a separator at the end with two particles. A desired construction is achieved in 57,400 generations and  an initial configuration of 56,240 cells. The whole evolution space is 3,228,176,000 cells. See details \cite{ctsr110}.

\subsection{Simulating a cyclic tag system in rule 110}
\label{simcts}

The cyclic tag system starts with the value `1' on the tape, see Fig.~\ref{CTS-diagrama}. We show a selection of snapshots of the machine working in rule 110 (see details in \cite{ctsr110, compressr110}). We show different sets of particles with coloured labels on the snapshot below.

\begin{figure}[th]
\centerline{\includegraphics[scale=.3]{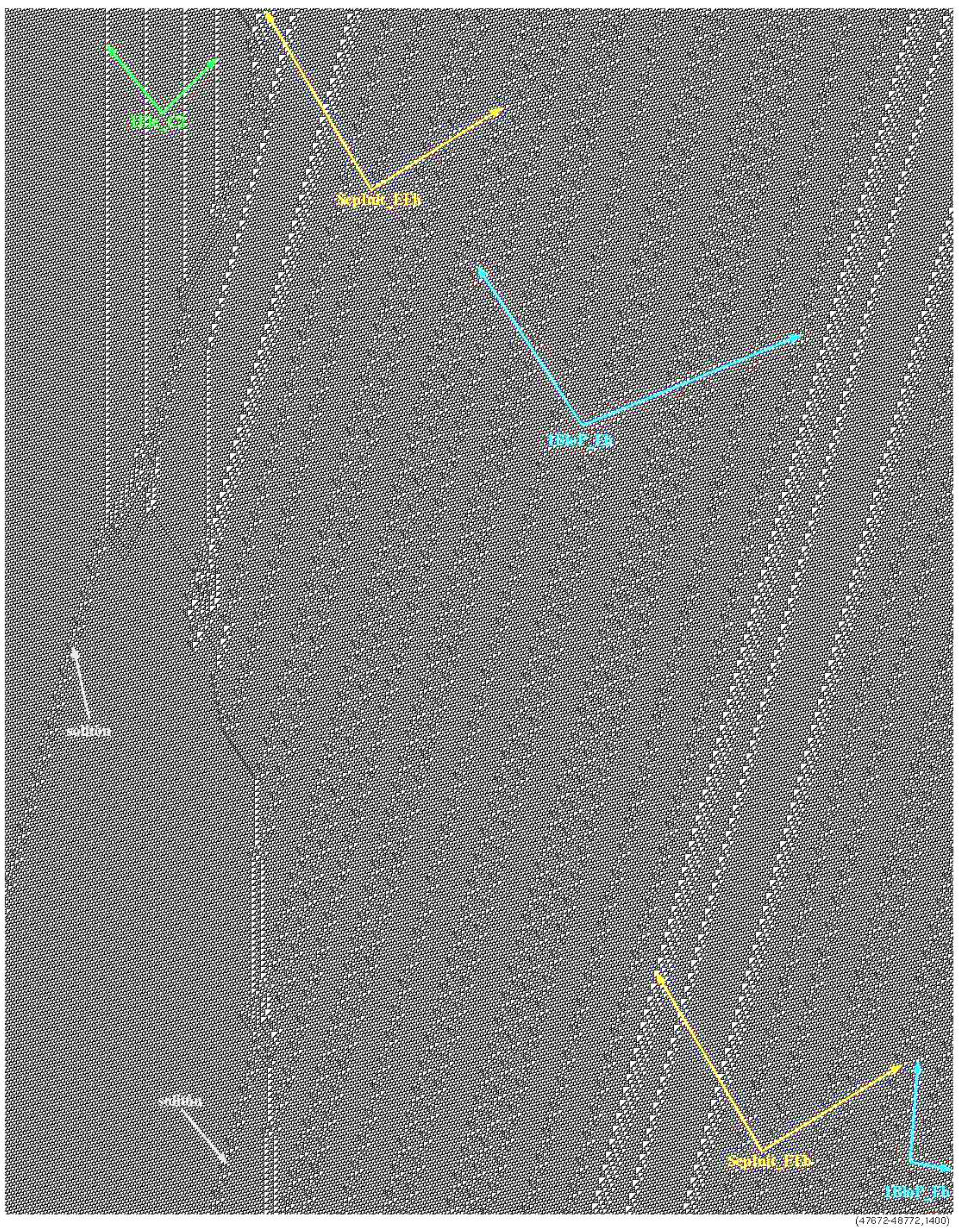}}
\caption{Initial stage of cyclic tag system in rule 110.}
\label{ctsCook-1}
\end{figure}

Figure~\ref{ctsCook-1} shows the initial stage of the cyclic tag system with the state `1' in the tape. This data is represented by the set of particles 1Ele$\_C_{2}$. The snaphshot shows a central part of the machine and a train of $A^{3}$ particles. We can see the first leader in the set of particles SepInit$\_E\bar{E}$ coming from the right periodic side. 

\begin{figure}[th]
\centerline{\includegraphics[scale=.3]{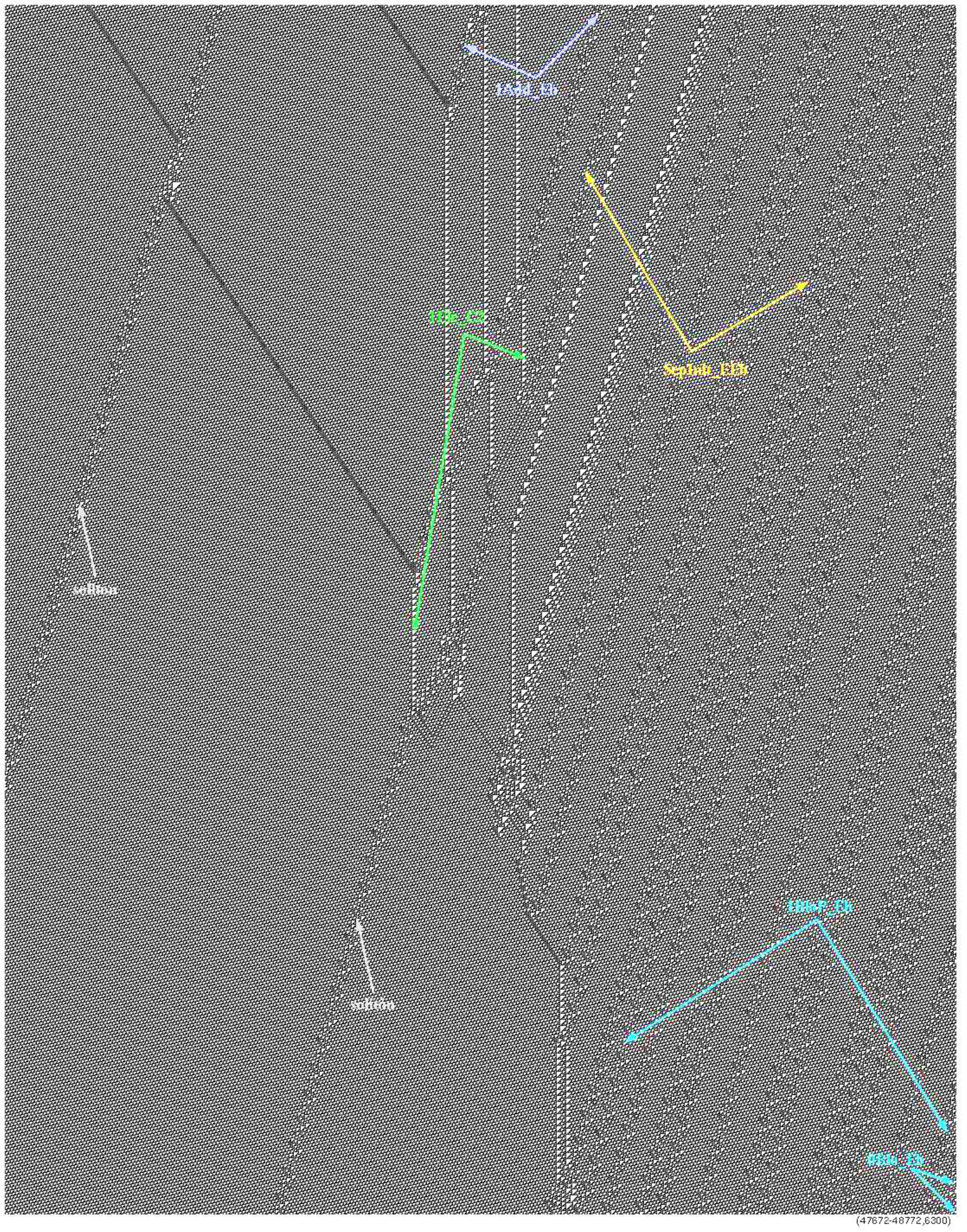}}
\caption{Constructing an element 1Ele$\_C_{2}$.}
\label{ctsCook-4}
\end{figure}

The first reaction in Fig.~\ref{ctsCook-1} deletes the state `1' on the tape. The set of particles 1Ele$\_C_{2}$) and the particles' separator are prepared for next data to be aggregated. If a set of particles 0Ele$\_C_{2}$ is encountered on the tape then data is not added to the tape until another separator appears. The particles $\bar{E}$ left after the first production are invisible to the system, they do not affect any operations because they cross as solitons, without state modifications, the subsequent set of particles 4$\_A^{4}$.

\begin{figure}[th]
\centerline{\includegraphics[scale=.3]{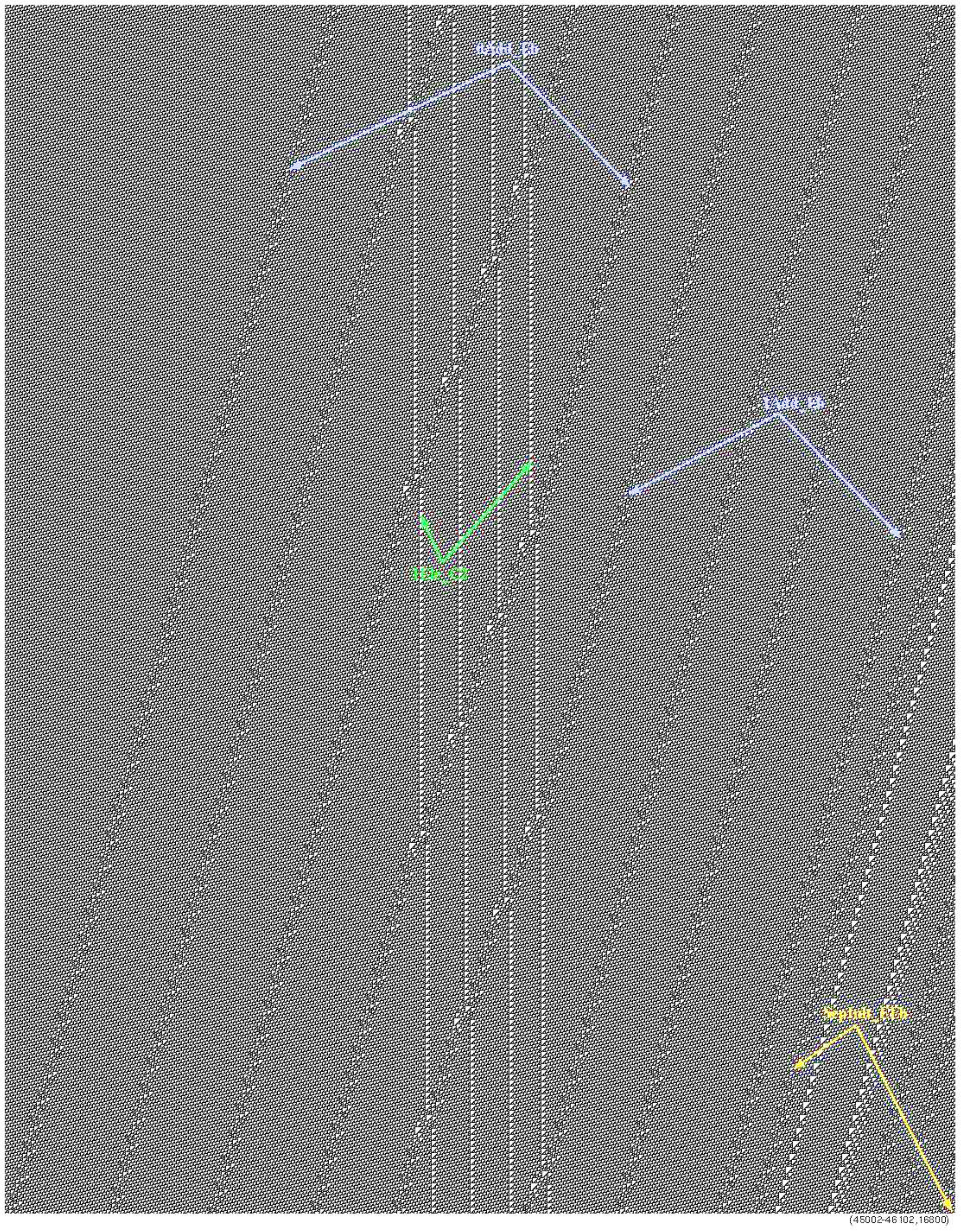}}
\caption{Transformed data crossing the tape of values.}
\label{ctsCook-7}
\end{figure}

In Fig.~\ref{ctsCook-4} we see a set of particles 1Ele$\_C_{2}$ constructed from a train of particles 4$\_A^{4}$. These particles have a very short life because quickly a separator set of particles arrives. This separator erases the particles and prepares new data that would be aggregated to the tape.

\begin{figure}[th]
\centerline{\includegraphics[scale=.3]{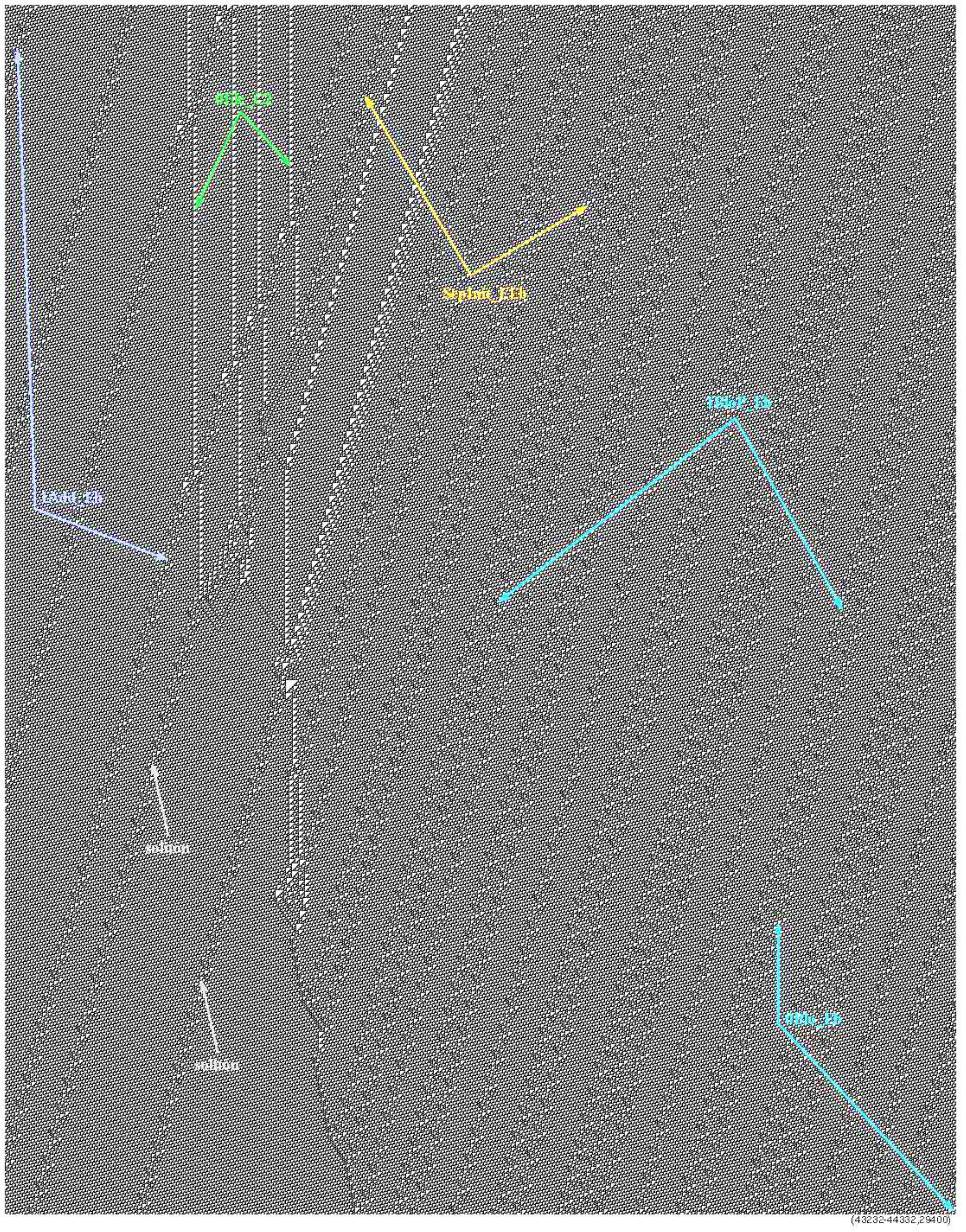}}
\caption{Deleting a set of particles 0Ele$\_$C2.}
\label{ctsCook-8}
\end{figure}

Figure~\ref{ctsCook-7} presents the construction of a set of particles 1Ele$\_C_{2}$. In this stage of the evolution, we can see how data is aggregated, based on their values, before they cross the tape. Similar reactions can be observed with the set of particles 0Ele$\_C_{2}$.

Figure~\ref{ctsCook-8} shows a constructed a set of particles 0Ele$\_C_{2}$ and its roles in the system. At the top, a set of particles 1Add$\_\bar{E}$, previously produced by a standard component 1BloS$\_\bar{E}$,  crosses a set of particles 0Ele$\_C_{2}$. A leader set of the particles deletes `0' from the tape and all the subsequent incoming data. There are 1BloP$\_\bar{E}$, 0Blo$\_\bar{E}$ and 1BloS$\_\bar{E}$ set of particles in the illustrated sequence. The tile $T_{14}$ is generated in the process. This differences in distances between the particles determine a change of phases which will lead to erasure of particles  $\bar{E}$, instead of production of particles $C$. The reaction $A^{3} \rightarrow \bar{E}$  is used to delete the particles.

Production rules in cyclic tag system specify that for the state `0' the first element of the chain must be erased and the other elements are conserved and no data are written on the tape. If the state  is `1' the first element of the chain is deleted and 10 or 11 are aggregated depending of the $k$ value. This behaviour is particularly visible when a separator finds 0 or 1 and deletes it from the tape. If the deleted data is `0', a separator does not allow the production of new data. If the deleted data is `1' the separator aggregates new elements 11 or 10, which are modified at later stages of the system's development. Using this procedure, we can calculate up to the sixth `1' of the sequence 011$<$1$>$0 produced by the cyclic tag system. 

In terms of periodic phases, this cyclic tag system working in rule 110 can be simplified as follows:

\begin{quotation}
\begin{tabbing}
{\bf left:} \{$649e$-4$\_A^{4}$(F$\_{i}$)\}*, for $1 \leq i \leq 3$ in sequential order \\ \\
{\bf center:} $246e$-1Ele$\_$C2(A,f$_{1}$\_1)-$e$-$A^{3}$(f$_{1}$\_1) \\ \\
{\bf right:} \= \{SepInit$\_E\bar{E}$(\#,f$_{i}$\_1)-1BloP$\_\bar{E}$(\#,f$_{i}$\_1)-SepInit$\_E\bar{E}$(\#,f$_{i}$\_1)- \\
\> 1BloP$\_\bar{E}$(\#,f$_{i}$\_1)-0Blo$\_\bar{E}$(\#,f$_{i}$\_1)-1BloS$\_\bar{E}$(\#,f$_{i}$\_1)\}* (where  \\
\> $1 \leq i \leq 4$ and \# represents a particular phase).
\end{tabbing}
\end{quotation}

These periodic coding will be very useful to design and synchronise three interlinked rings of 1D CA (cyclotrons) to make a `supercollider'.

\section{Cellular automata supercollider}
\label{supercollider}

In the late 1970s Fredkin and Toffoli proposed a concept of  computation based on ballistic interactions between quanta of information that are represented by abstract particles~\cite{supercolliders}. The Boolean states of logical variables are represented by balls or atoms, which preserve their identity when they collide with each other.  Fredkin, Toffoli and Margolus developed a billiard-ball model of computation, with underpinning mechanics of elastically colliding balls and mirrors reflecting the balls' trajectories. Margolus proposed a special class of CA which implements the billiard-ball model~\cite{physcomp}. Margolus' partitioned CA exhibited computational universality because they simulated Fredkin gates via collision of soft spheres~\cite{softspheres, crystcomp}. Also, we consider previous results about circular machines designed by Arbib, Kudlek, and Rogozhin in \cite{arbibbook, newpostm, smallpostm}. Initial reports about CA collider were published in \cite{cacollider, comprings, saicollider}.

\begin{figure}[th]
\sidecaption
\subfigure[]{\includegraphics[scale=.45]{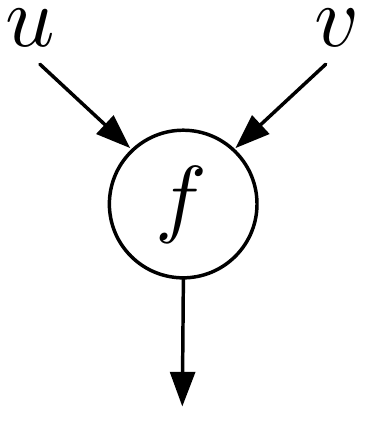}} \hspace{1cm}
\subfigure[]{\includegraphics[scale=.45]{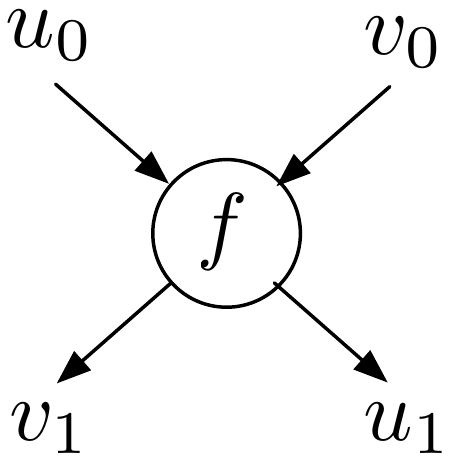}} \hspace{1cm}
\subfigure[]{\includegraphics[scale=.45]{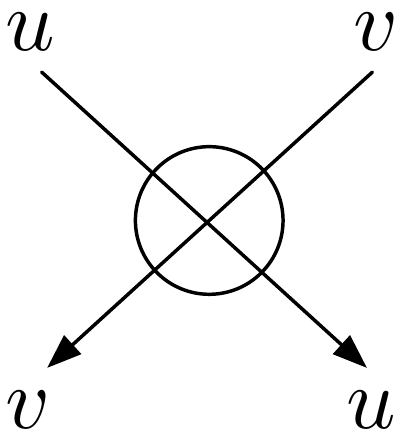}}
\hspace{1cm}
\subfigure[]{\includegraphics[scale=.45]{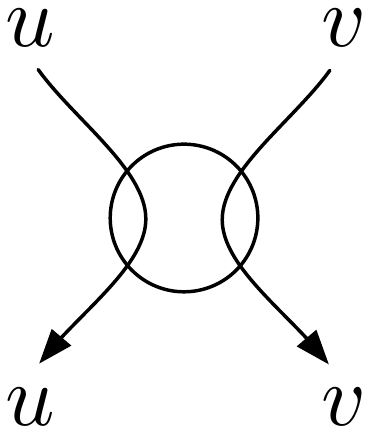}}
\caption{Schemes of ballistic collision between localizations representing logical values of the Boolean variables $u$ and $v$.}
\label{particles}
\end{figure}

The following functions with two input arguments $u$ and $v$ can be realised in collisions between two localizations:

\begin{quotation}
\begin{itemize}
\item $f(u,v) = c$, fusion (Fig.~\ref{particles}a)
\item $f(u,v) = u+v$, interaction and subsequent change of state (Fig.~\ref{particles}b)
\item $f_i(u,v) \mapsto (u,v)$ identity, solitonic collision (Fig.~\ref{particles}c);
\item $f_r(u,v) \mapsto (v,u)$ reflection, elastic collision (Fig.~\ref{particles}d);
\end{itemize}
\end{quotation}

To represent Toffoli's supercollider~\cite{supercolliders} in 1D CA we use the notion of an idealised particle $p \in {\texttt G}$ (without energy and potential). The particle $p$ is represented by a binary string of cell states.

\begin{figure}[th]
\sidecaption
\subfigure[]{\includegraphics[scale=.5]{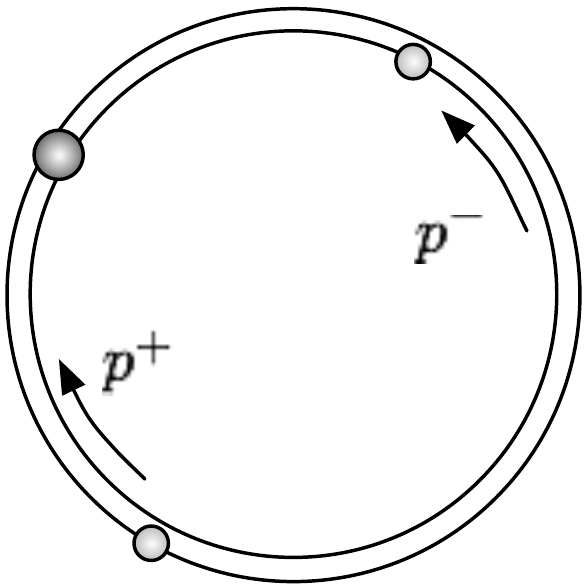}} \hspace{0.8cm}
\subfigure[]{\includegraphics[scale=.5]{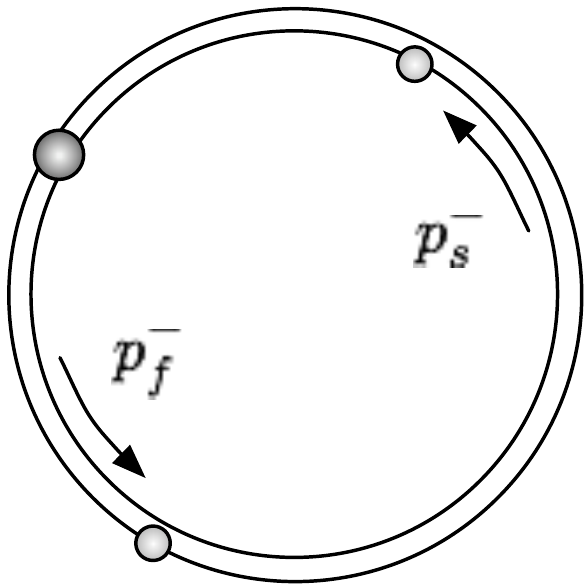}} \hspace{0.8cm}
\subfigure[]{\includegraphics[scale=.5]{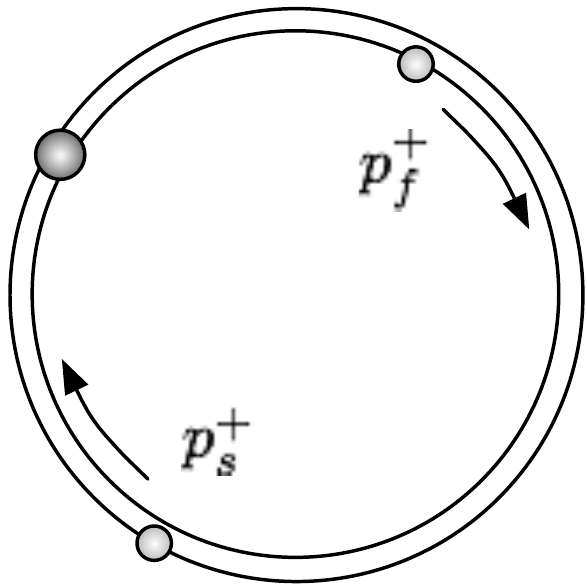}}
\caption{Representation of abstract particles in a 1D CA  ring.}
\label{beamRouting}
\end{figure}

Figure~\ref{beamRouting} shows two typical scenarios where particles $p_f$ and $p_s$ travel in a CA  cyclotron. The first scenario (Fig.~\ref{beamRouting}a) shows two particles travelling in opposite directions; these particles collide one with another. Their collision site (contact point) is shown by a dark circle in Fig.~\ref{beamRouting}a. The second scenario demonstrates a beam routing where a fast particle $p_f$ eventually catches up with a slow particle $p_s$ at a collision site (Fig.~\ref{beamRouting}b). If the particles collide like solitons, then the faster particle $p_f$ simply overtakes the slower particle $p_s$ and continues its motion (Fig.~\ref{beamRouting}c).

\begin{figure}[th]
\centerline{\includegraphics[scale=.43]{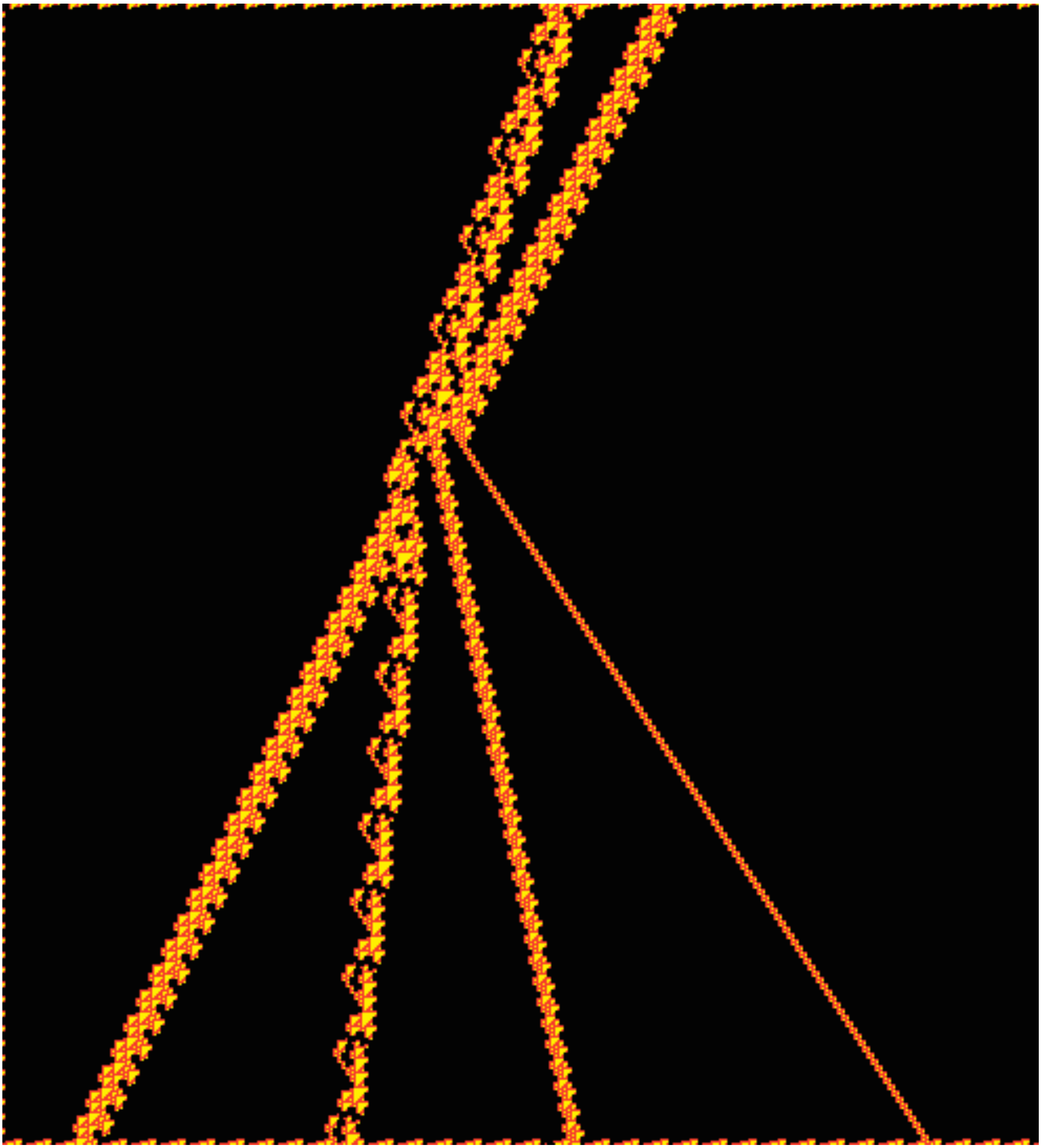}}
\caption{Particle collision in rule 110. Particle $p^-_{\bar{B}}$ collides with particle $p^-_G$ giving rise to three new particles --- $p^-_{F}$, $p^+_{D_2}$, and $p^+_{A^3}$, and preserving the $p^-_{\bar{B}}$ particle --- that are generated as a result of the collision.}
\label{splitParticleR110}
\end{figure}

Typically, we can find all types of particles in complex CA, including particles with positive $p^+$, negative $p^-$, and neutral $p^0$ displacements, and composite particles assembled from elementary localizations. A sample coding and colliding particles is shown in Fig.~\ref{splitParticleR110}, which displays a typical collision between two particles in rule 110. As a result of the collision one particle is split into three different particles (for full details please see \cite{atlasr110}). The previous collision positions of particles determines the outcomes of the collision. Particles are represented now with orientation and name of the particle in rule 110 as follows: $p^{+,-,0}_{\texttt G}$.

\begin{figure}[th]
\sidecaption
\subfigure[]{\includegraphics[scale=.23]{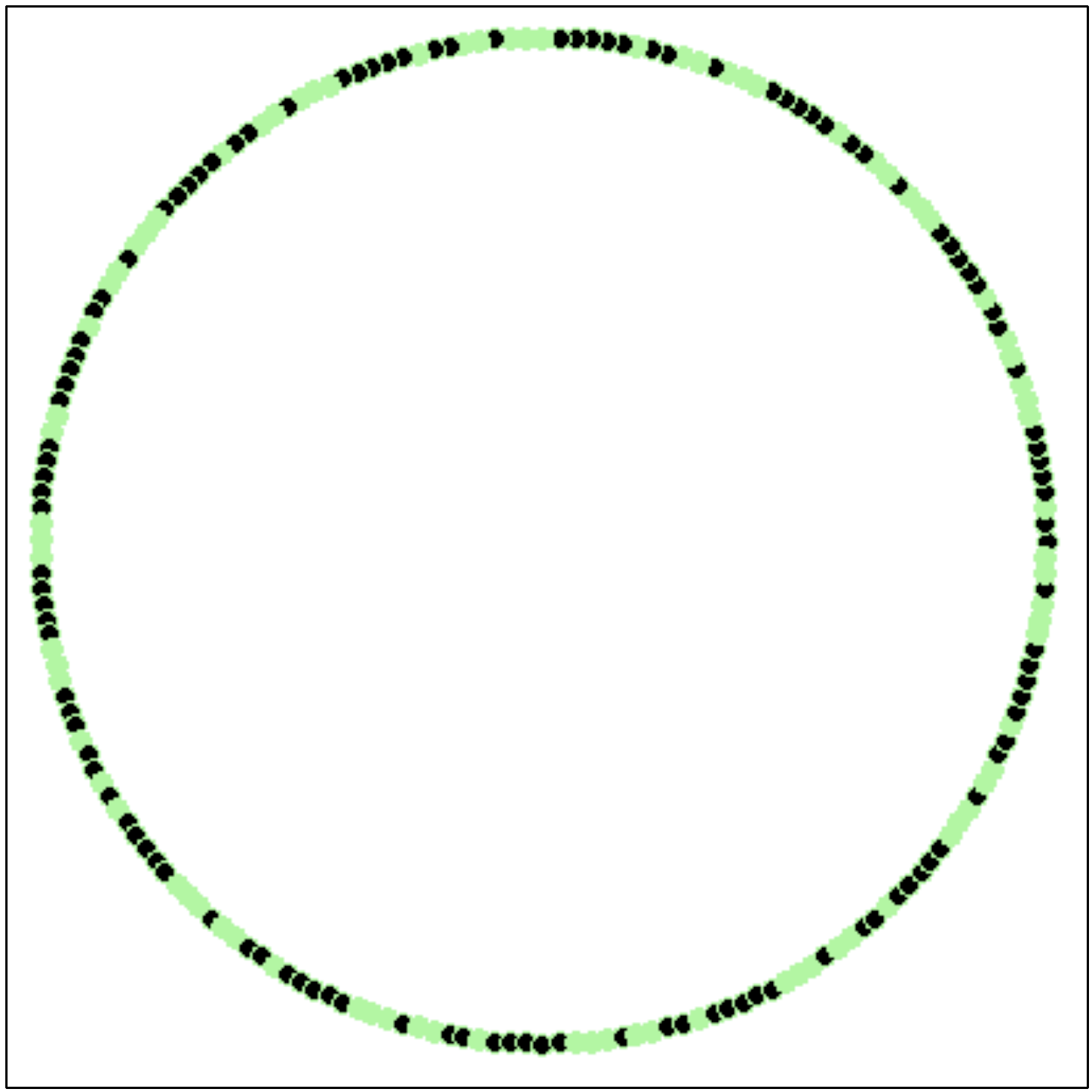}} \hspace{0.2cm}
\subfigure[]{\includegraphics[scale=.235]{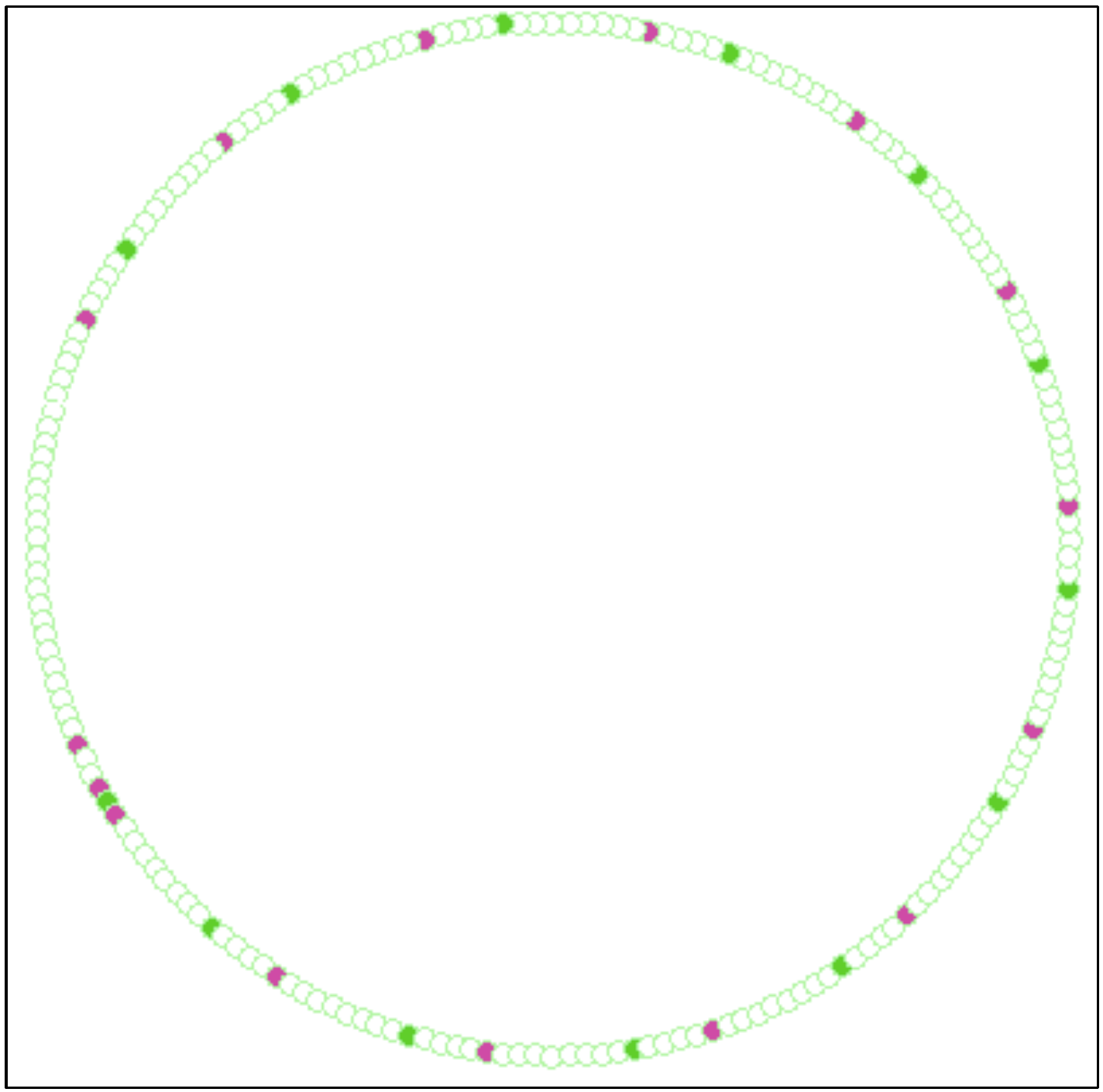}} \hspace{0.2cm}
\subfigure[]{\includegraphics[scale=.455]{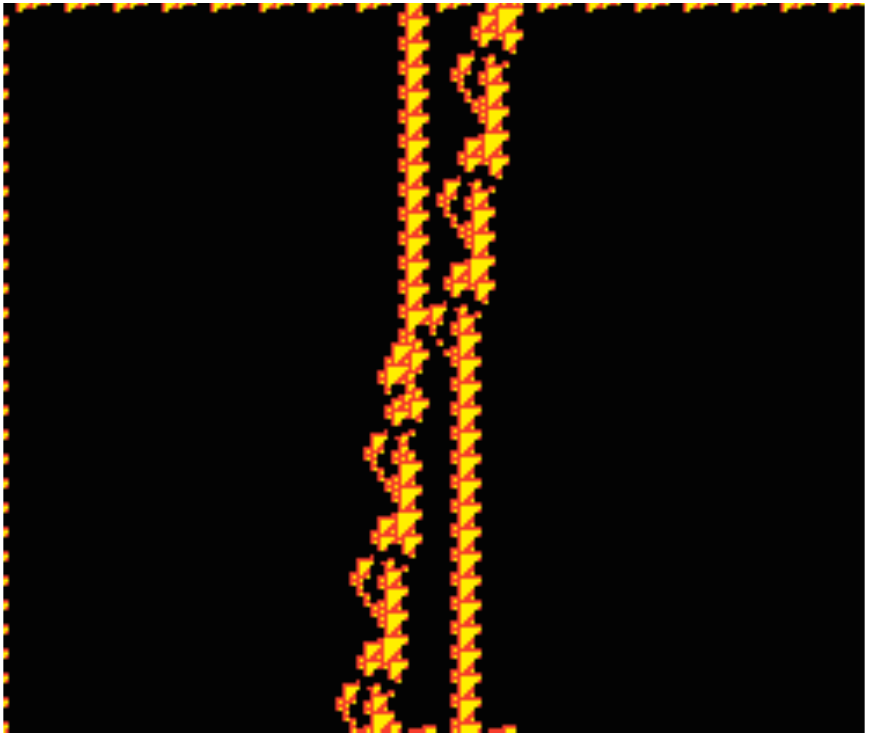}} 
\caption{A soliton-type interaction between particles in rule 110: (a)--(b)~two steps of beam routing, (c)~exact configuration at the time of collision.}
\label{solitonR110}
\end{figure}

To represent particles on a given beam routing scheme (see Fig.~\ref{beamRouting}), we do not consider the periodic background configuration in rule 110 because essentially this does not affect on collisions. Fig.~\ref{solitonR110} displays a 1D configuration where two particles collide repeatedly and interact as solitons so that the identities of the particles are preserved in the collisions. A negative particle $p^-_F$ collides with and overtakes a neutral particle $p^-_{C_1}$. First cyclotron (Fig.~\ref{solitonR110}a) presents a whole set of cells in state 1 (dark points) evolving with the periodic background. By applying a filter we can see better these interactions (Fig.~\ref{solitonR110}b).\footnote{Cyclotron evolution was simulated with DDLab software, available at \url{http://www.ddlab.org}.} Typical space-time configurations of a CA exhibiting a collision between $p^-_F$ and $p^-_{C_1}$ particles are shown in Fig.~\ref{solitonR110}c.

\section{Beam routings and computations}
\label{routing}

\begin{figure}[th]
\centerline{\includegraphics[scale=.5]{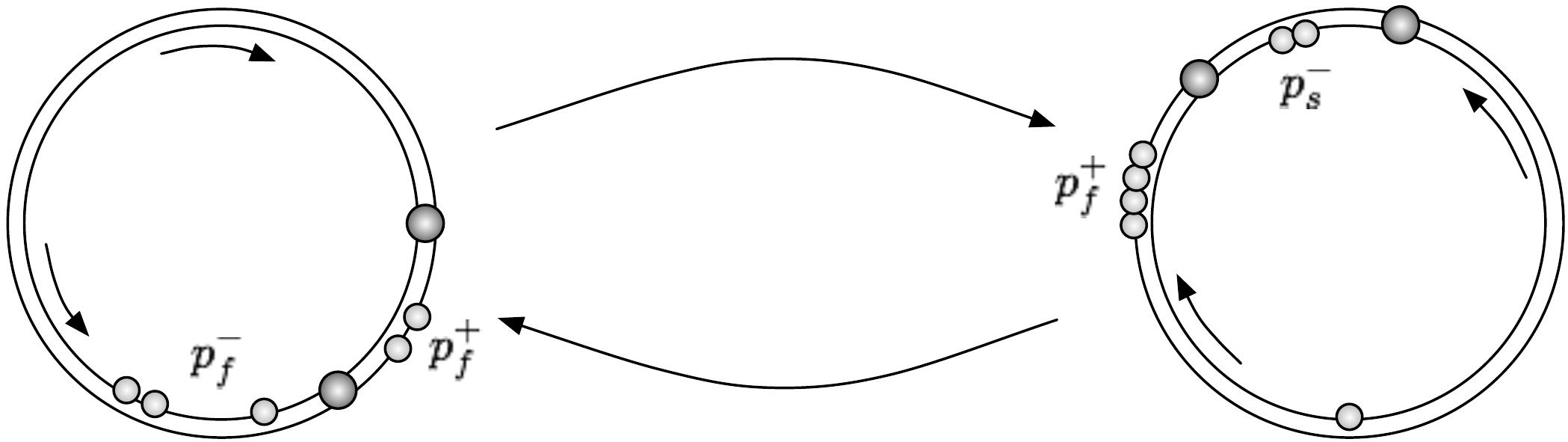}}
\caption{Transition between two beam routing synchronising multiple reactions. When the first set of collisions is done a new beam routing is defined with other set of particles, so that when the second set of collisions is done then first beam returns to its original state.}
\label{beamRoutingTransition}
\end{figure}

We examine beam routing based on particle-collisions. We will show how the beam routing can be used in designs of computing based-collisions connecting cyclotrons. Figure~\ref{beamRoutingTransition} shows a beam routing \index{beam routing} design, connecting two of beams and then creating a new beam routing diagram where edges represent a change of particles and collisions. In such a transition, new particles emerge and collide to return to the first beam. The particles oscillate between these two beam routing indefinitely.

\begin{figure}[th]
\centerline{\includegraphics[scale=.52]{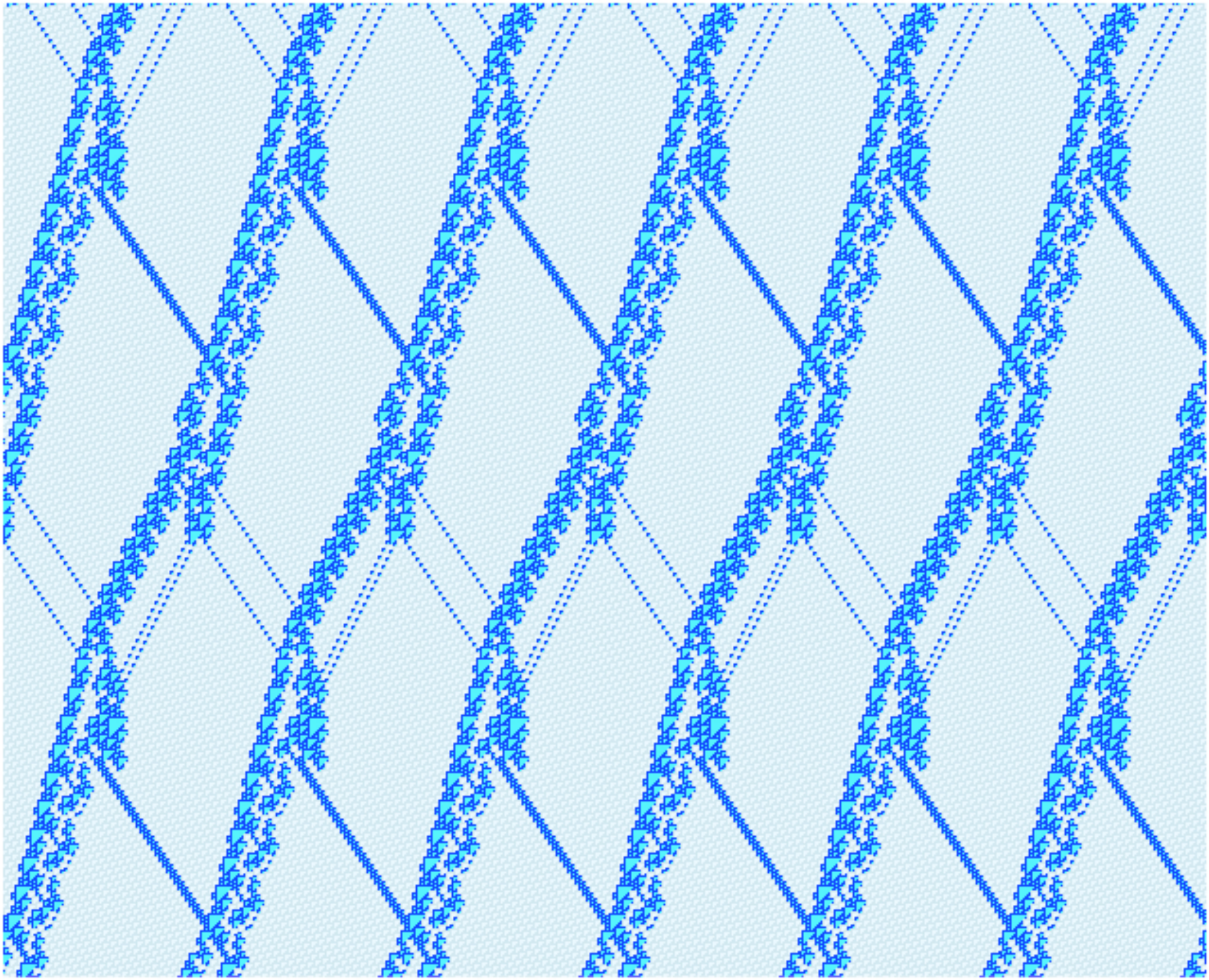}}
\caption{Synchronisation of multiple collisions in rule 110 on a ring of 1,060 cells in 1,027 generations, starting with 50 particles from its initial condition.}
\label{metaR110}
\end{figure}

To understand how dynamics of a double beam differs from a conventional 1D evolution space we provide Fig.~\ref{metaR110}. There we can see  multiple collisions between particles from first beam routing and trains particles. Exactly, we have that

$$
p_{A}^+, p_{A}^+ \leftrightarrow p_{\bar{B}}^-, p_{B}^-, p_{B}^-
$$

\noindent changes to the set of particles derived in the second beam routing:

$$
p_{A^4}^+ \leftrightarrow p_{E}^+, p_{\bar{E}}^+.
$$

This oscillation determines two beam routing connected by a transition of collisions as:

$$
(p_{A}^+, p_{A}^+ \leftrightarrow p_{\bar{B}}^-, p_{B}^-, p_{B}^-) \rightarrow (p_{A^4}^+ \leftrightarrow p_{E}^+, p_{\bar{E}}^+) \mbox{, and }$$
$$
(p_{A^4}^+ \leftrightarrow p_{E}^+, p_{\bar{E}}^+) \rightarrow (p_{A}^+, p_{A}^+ \leftrightarrow p_{\bar{B}}^-, p_{B}^-, p_{B}^-).
$$

We can see that a beam routing representation allows for a design of collisions in cyclotrons. We employ the beam routing to implement the cyclic tag system in the CA rings. A construction of the cyclic tag system in rule 110 consists of three components  (as was discussed in Sect.~\ref{simcts}):

\begin{itemize}
\item The {\it left periodic part}, controlled by trains of 4\_$A^{4}$ particles. This part is static. It controls the production of $0$'s and $1$'s. 
\item The {\it centre}, determining the initial value in the tape.
\item The {\it right periodic part}, which has the data to process, adding a leader component which determines if data will be added or erased in the tape.
\end{itemize}

\begin{figure}[th]
\sidecaption
\subfigure[]{\scalebox{0.6}{\includegraphics{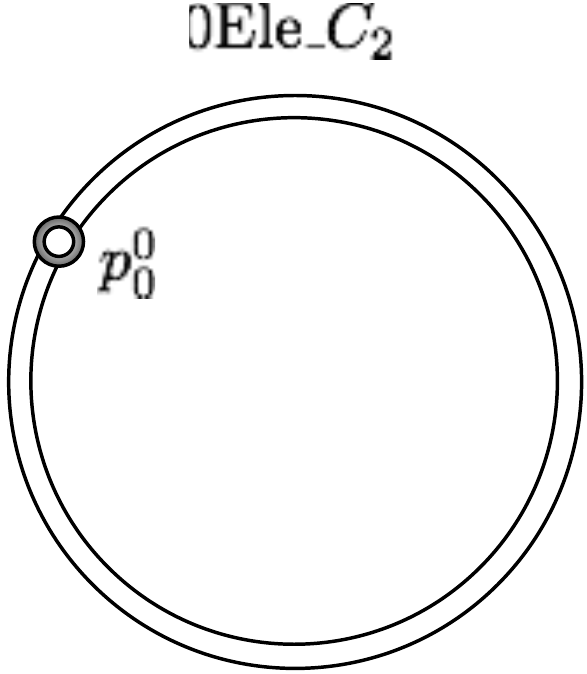}}} \hspace{0.1cm}
\subfigure[]{\scalebox{0.6}{\includegraphics{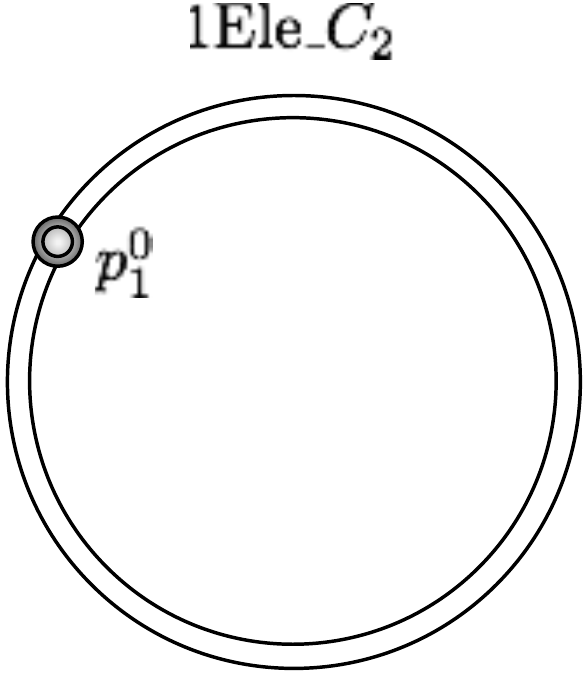}}} \hspace{0.1cm}
\subfigure[]{\scalebox{0.6}{\includegraphics{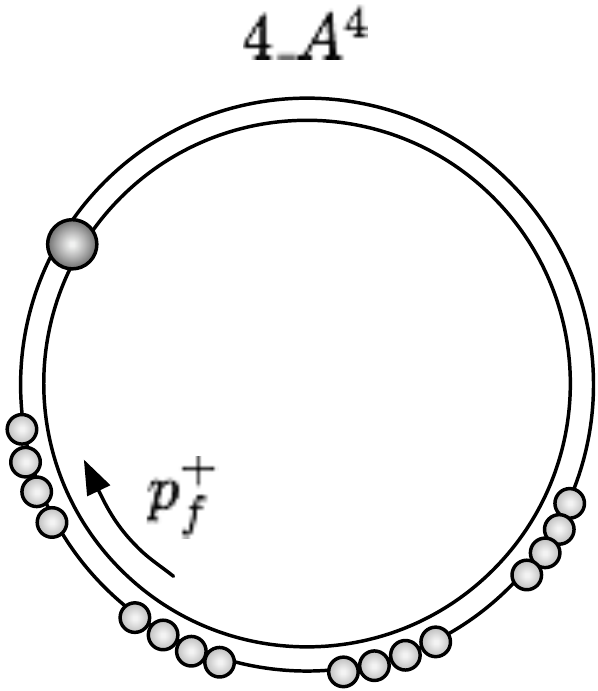}}} \hspace{0.1cm}
\subfigure[]{\scalebox{0.6}{\includegraphics{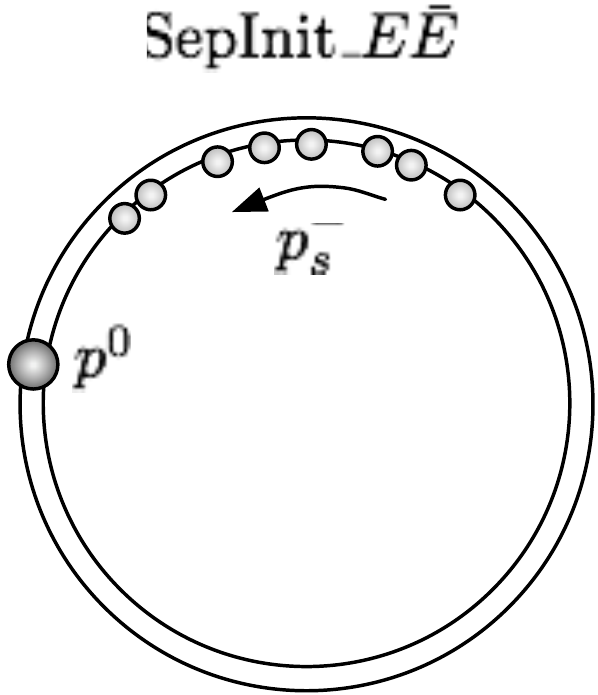}}} \hspace{0.1cm}
\subfigure[]{\scalebox{0.6}{\includegraphics{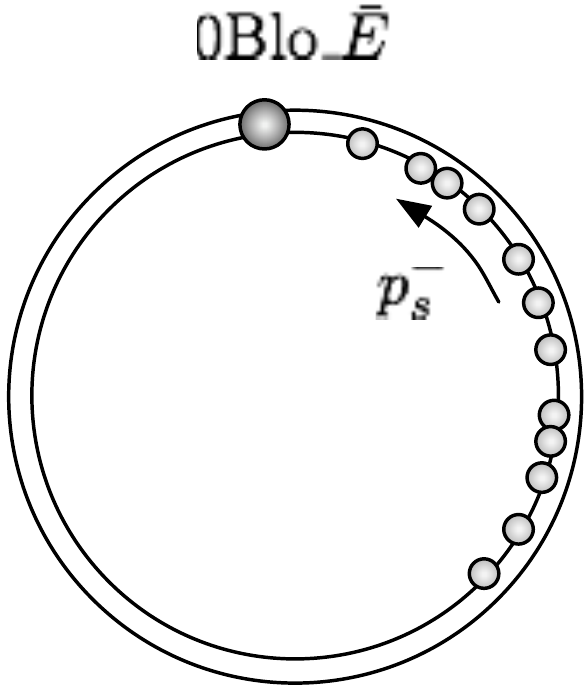}}} \hspace{0.1cm}
\subfigure[]{\scalebox{0.6}{\includegraphics{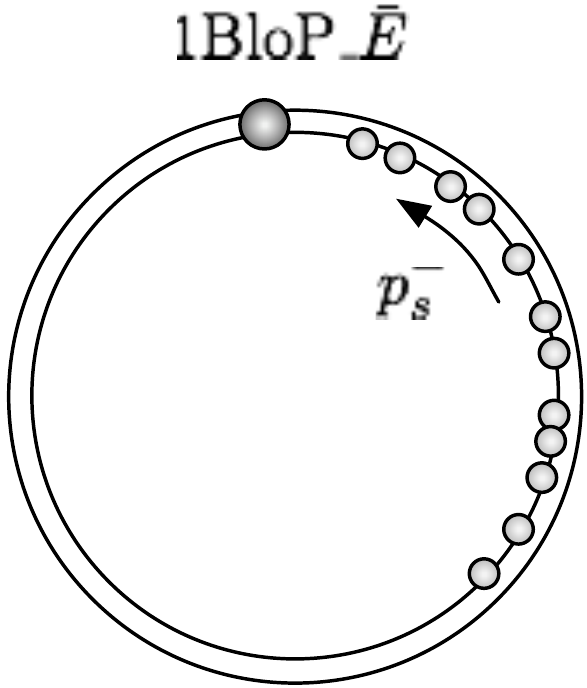}}} \hspace{0.1cm}
\subfigure[]{\scalebox{0.6}{\includegraphics{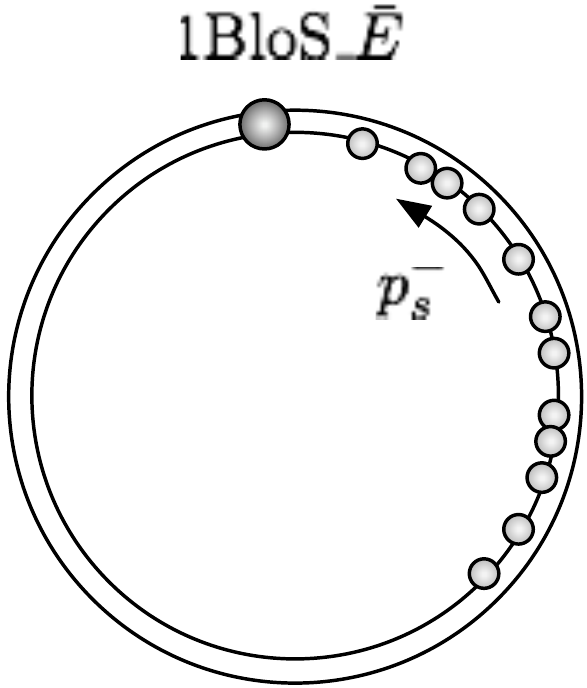}}} \hspace{0.1cm}
\subfigure[]{\scalebox{0.6}{\includegraphics{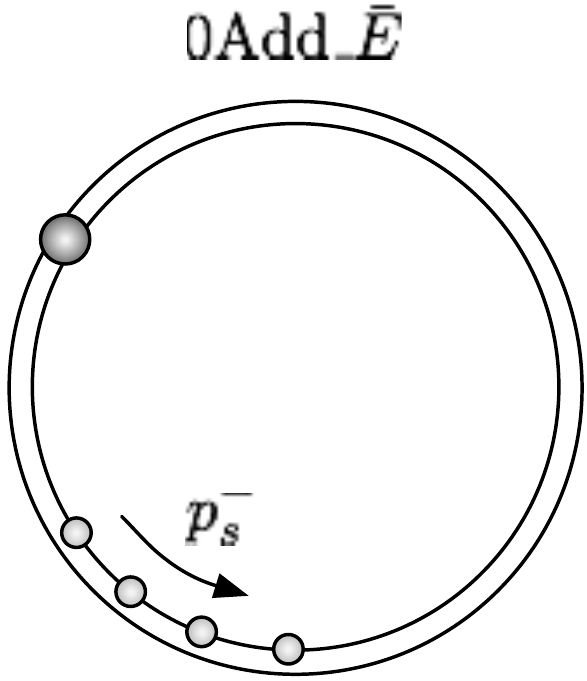}}} \hspace{0.1cm}
\subfigure[]{\scalebox{0.6}{\includegraphics{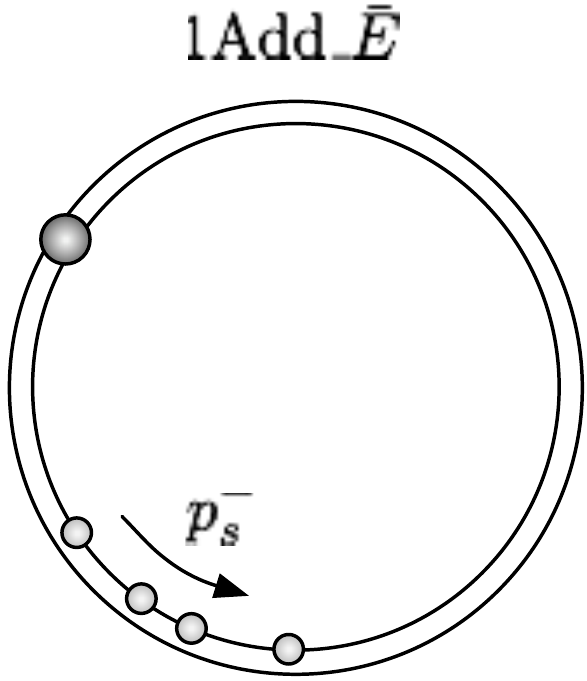}}} 
\caption{The whole set of beam routing codification representing train of particles, to simulate a cyclic tag system. Each global state represents every component (set of particles) described in Section~\ref{compcts}.}
\label{beamRoutingR110}
\end{figure}

\begin{figure}
\centerline{\includegraphics[width=4in]{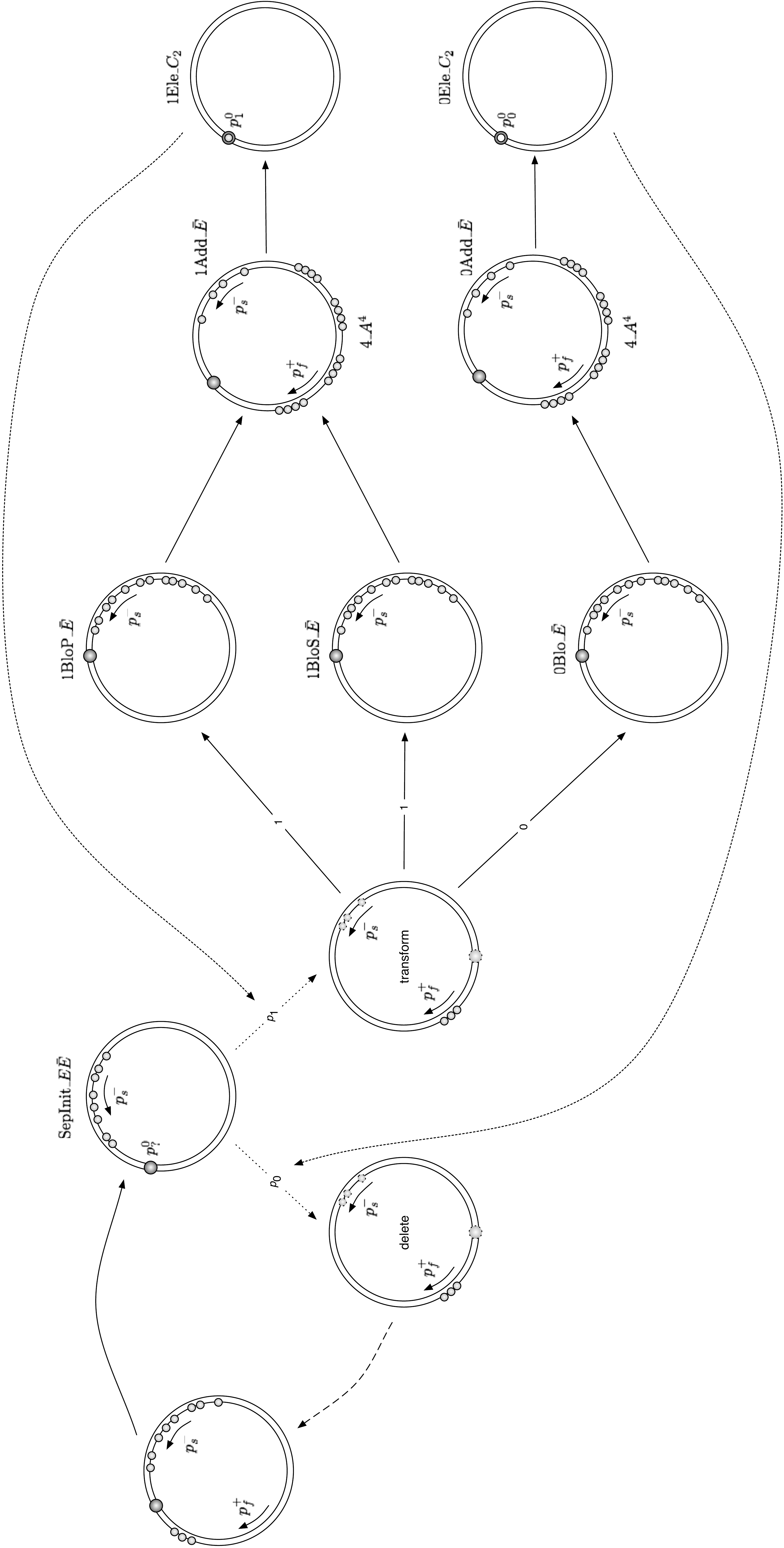}}
\caption{Beam routing finite state machine simulating the cyclic tag system by state of cyclotrons representation.}
\label{CTSbeamrouting}
\end{figure}

{\it Left periodic part} is defined by four trains of $A^{4}$ (Fig.~\ref{beamRoutingR110}c), trains of $A^{4}$ have three phases. The key point is to implement these components defining both distances and phases, because a distinct phase or a distance induces an undesirable reaction.

The {\it central part} is represented by one value `1' on the tape across a train of four $C_{2}$ particles. The component 1Ele$\_C_{2}$ (Fig.~\ref{beamRoutingR110}b) represents `1' and the component 0Ele$\_C_{2}$ (Fig.~\ref{beamRoutingR110}a) represents `0' on the tape. The component 0Blo$\_\bar{E}$ is formed by 12$\bar{E}$ particles. The construct includes two components to represent the state `1':  1BloP$\_\bar{E}$ (Fig.~\ref{beamRoutingR110}f) named {\it primary} and 1BloS$\_\bar{E}$ (Fig.~\ref{beamRoutingR110}g) named {\it standard}. A leader component SepInit$\_E\bar{E}$ (Fig.~\ref{beamRoutingR110}d) is used to separate trains of data and to determine their incorporation into of the tape.

The components 1Add$\_\bar{E}$ (Fig.~\ref{beamRoutingR110}i) and 0Add$\_\bar{E}$ (Fig.~\ref{beamRoutingR110}h) are produced by two previous different trains of data. The component 1Add$\_\bar{E}$ must be generated by a block 1BloP$\_\bar{E}$ or by 1BloS$\_\bar{E}$. This way, both components can yield the same element. The component 0Add$\_\bar{E}$ is generated by a component 0Blo$\_\bar{E}$ (Fig.~\ref{beamRoutingR110}e). For a complete and full description of such reproduction by phases f$_{i}\_1$, see \cite{ctsr110}.

To get a cyclic tag system emulation in rule 110 by beam routings, we will use connections between beam routings as a finite state machine represented in Fig~\ref{CTSbeamrouting}. Transitions between beam routings means a change of state (transition function). Initial state is represented by the component 1Ele$\_C_2$. A final state is not specified because it is determined by the state of the computation, i.e., a halt condition. Components 1Ele$\_C_2$ and 0Ele$\_C_2$ are compressed and shown as a dark circle, which represents the point of collision. Both components are made of four $C_2$ particles being at different distances. When a leader component (SepInit$\_E\bar{E}$) is transformed, given previous binary value on the tape, it collides with $p_?^0$ component, i.e., a $p_1^0$ or $p_0^0$ element. If $p_?^0$ is `0', then a cascade of collisions starts to {\it delete} all components that come with three particles successively. If $p_?^0$ is `1' then a cascade of {\it transformations} dominated by additional particles $p^0$ is initiated, in order to reach the next leader component. Here, we have more variants because pre-transformed train of particles is encoded into binary values that are then written on the machine tape. If a component of particles is 1BloP$\_\bar{E}$ or 1BloS$\_\bar{E}$ this means that such a component will be transformed to one 1Add$\_\bar{E}$ element.  If a component of particles is 0Blo$\_\bar{E}$, then such a component will be transformed to 0Add$\_\bar{E}$ element. At this stage, when both components are prepared then a binary value is introduced on the tape, a 1Add$\_\bar{E}$ element stores a 1 (1Ele$\_C_2$), and a 0Add$\_\bar{E}$ element stores a 0 (0Ele$\_C_2$), which eventually will be deleted for the next leader component and starts a new cycle in the cyclic tag system. In bigger spaces these components will be represented just as a point in the evolution space: we describe this representation in the next section.

\section{Cyclotrons}
\label{cyclotrons}

\begin{figure}[th]
\centerline{\includegraphics[scale=.342]{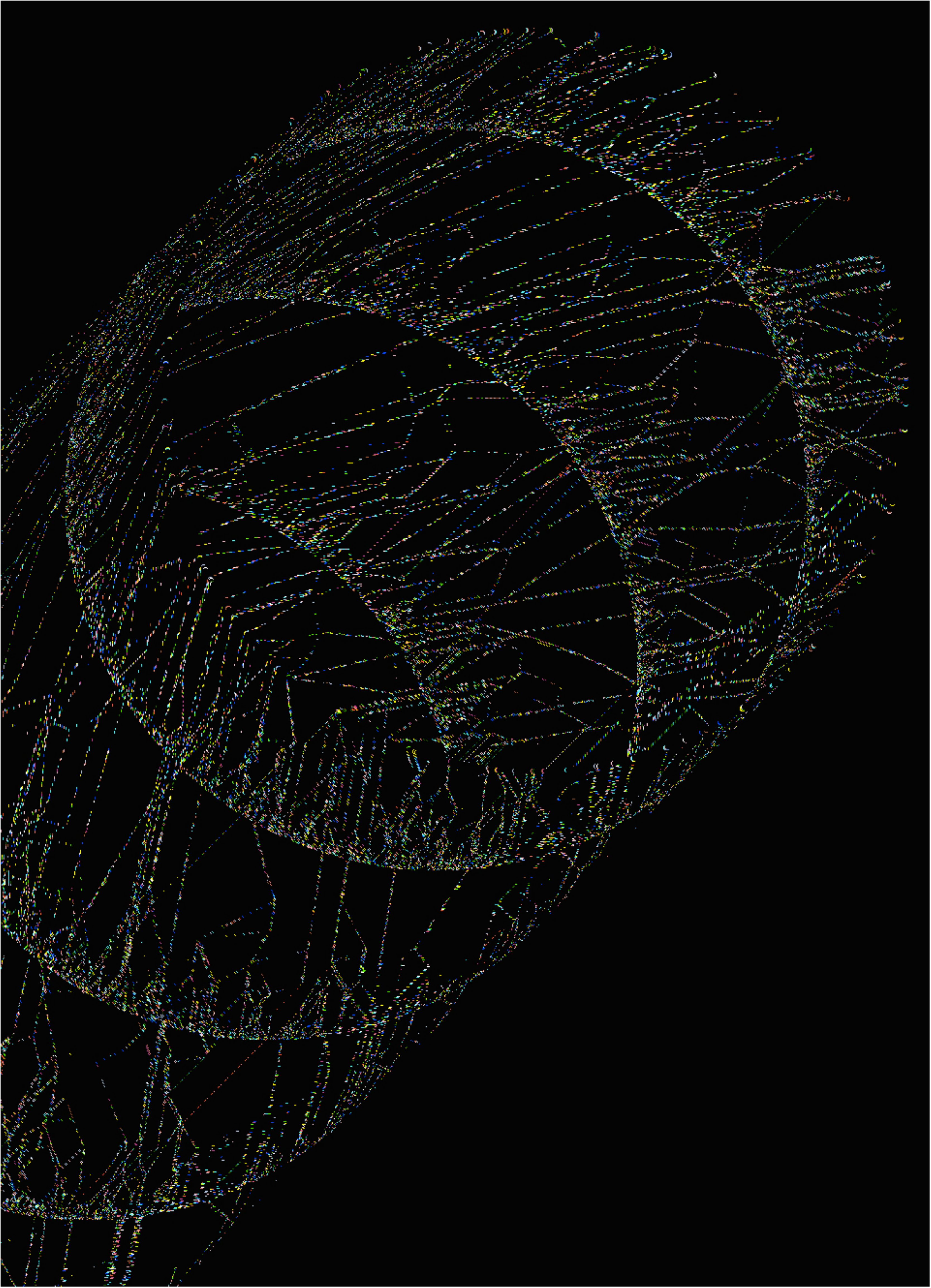}}
\caption{ECA rule 110 particles traveling and colliding inside a cyclotron in a evolution space of 20,000 cells. A filter is selected for a better view of particles, each cyclotron initial stage in the history (three dimensional projection) is restarted randomly to illustrate the complex dynamics and variety of particles and collisions. }
\label{r110_20000_rand}
\end{figure}

We use cyclotrons to explore large computational spaces where exact structures of particles are not relevant but only the interactions between the particles. There we can represent the particles as points and trains of particles as sequences of points. A 3D representation is convenient to understand the history of the evolutions, number of particles, positions, and collisions. Fig.~\ref{r110_20000_rand} shows a cyclotron evolving from a random initial configuration with 20,000 cells. Three stages are initialised in the evolution and the particles undergo successions of collisions in few first steps of evolution. The evolution is presented in a vertical orientation rotated 90 degrees. The present state shown is a front and its projection in three dimensions unveils the history and the evolution. Following this representation we can design a number of initial conditions to reproduce periodic patterns.\footnote{The simulations are done in {\it Discrete Dynamics Lab} (DDLab, \url{http://www.ddlab.org/}) \cite{ddlabbook}.}  

\begin{figure}[th]
\centerline{\includegraphics[scale=.344]{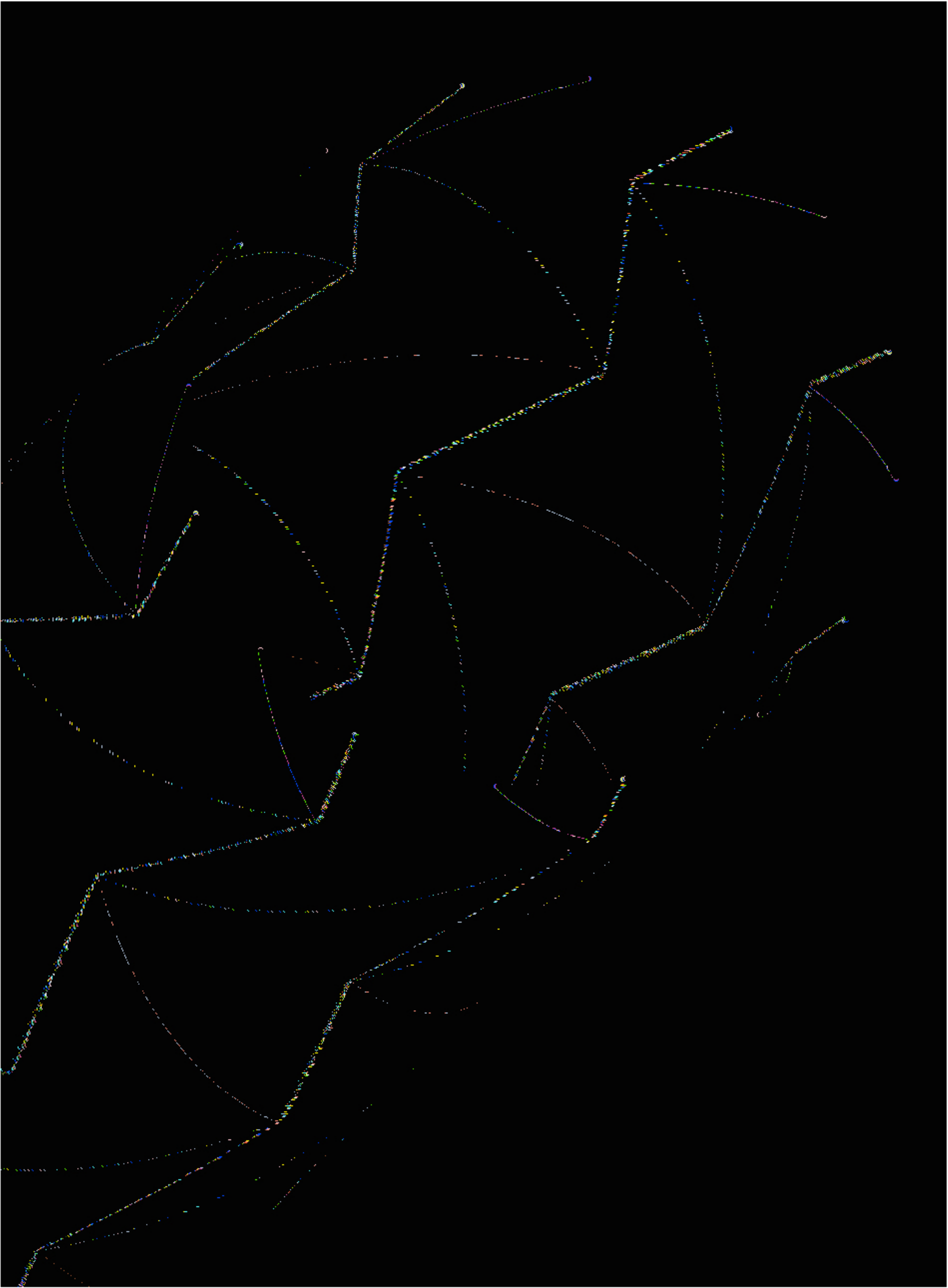}}
\caption{Basic flip-flop oscillator implemented in a cyclotron with 7,464 cells in 25,000 generations. 16 particles $p_{F} \leftarrow p_{B}$ were coded.}
\label{r110meta7464-3}
\end{figure}

Figure~\ref{r110meta7464-3} shows a basic flip-flop pattern. We synchronise 16 particles $p_{F} \leftarrow p_{B}$, the basic collision takes place for two pairs of particles, a $p_{D_1}$ particle and a train of $p_{A^2}$ particles. The distance is determined by a factor of {\sf mod} 14. A second reaction is synchronised with $p_{D_1} \leftarrow p_{A^2}$ to return back to the initial $p_{F}$ and $p_{B}$ particles. All 16 particles are forced in the same phase to guarantee an adequate distance, this distance is fixed in 64 copies of 14 cells (ether). Finally eight collisions are controlled every time simultaneously on an evolution space with 7,464 cells.

\newpage

\newpage

\newpage

\section{Collider computing}
\label{collidercomp}

\begin{figure}[th]
\centerline{\includegraphics[scale=.47]{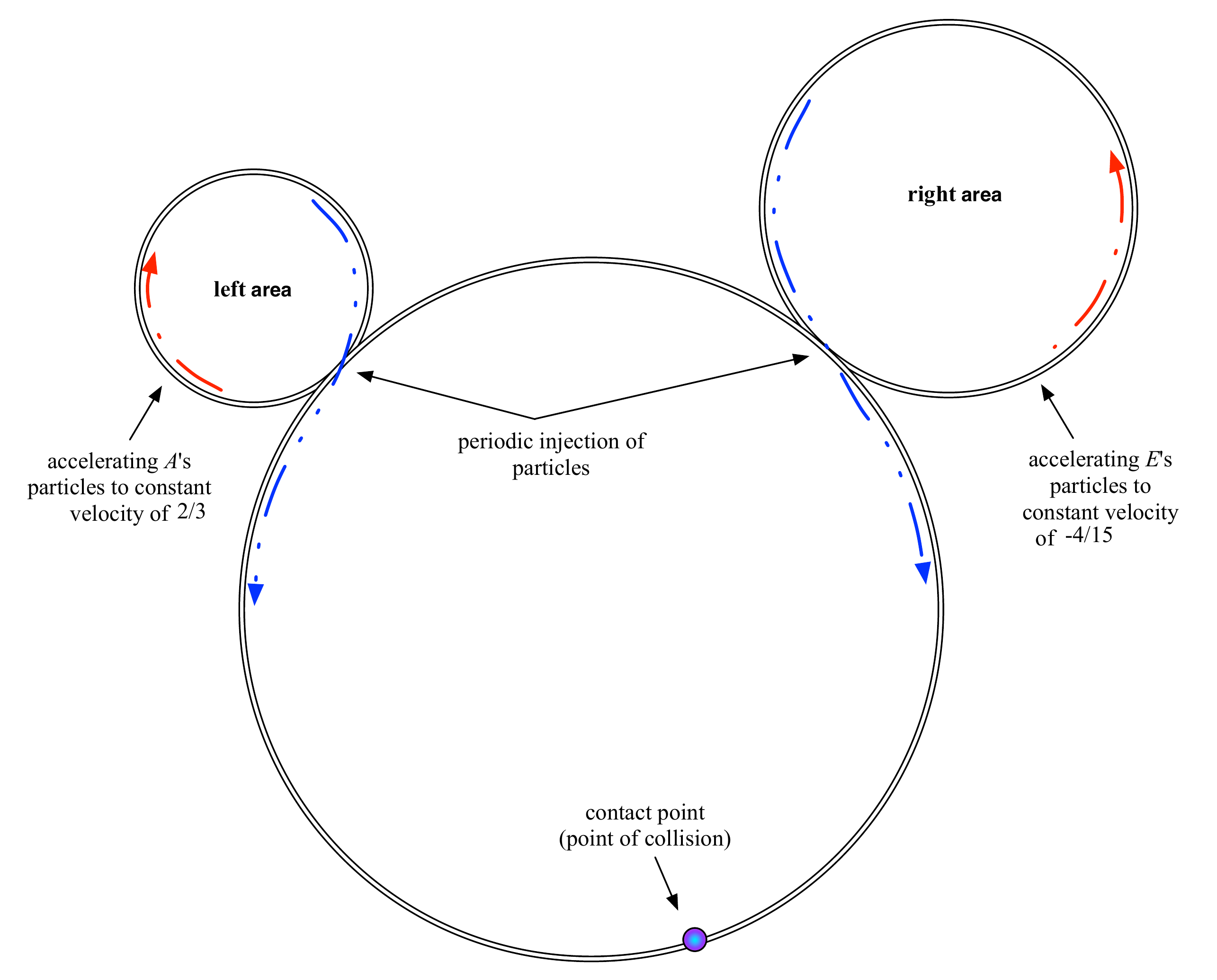}}
\caption{Collider diagram.}
\label{colliderDiagram}
\end{figure}

A cyclic tag system consists of three main components. Each stage of computation can be represented with a cyclotron. A synchronisation of these cyclotrons injects beams of particles to a central main collider to obtain the collisions that will simulate a computation. The periodic representations of left and right cyclotrons are fixed. Diagram in Fig.~\ref{colliderDiagram} shows the dynamics of particles in a collider.

\begin{figure}[th]
\centerline{\includegraphics[scale=.34]{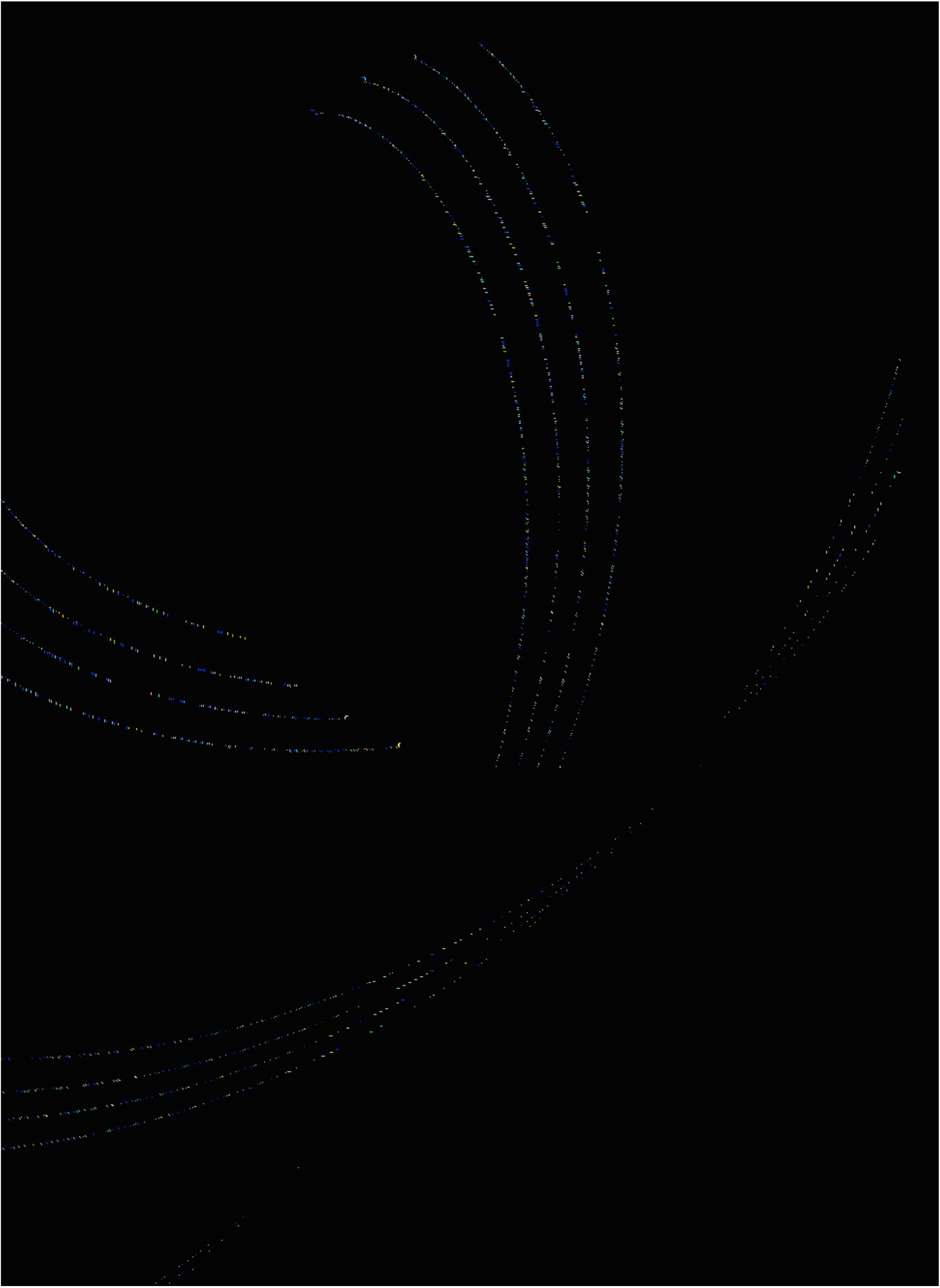}}
\caption{Three beams of $4p_{A^4(F_{i})}$ particles. Simulation is displayed in a vertical position to get a better view of particles' trajectories.}
\label{CTSleftArea}
\end{figure}

\begin{figure}[th]
\centerline{\includegraphics[scale=.34]{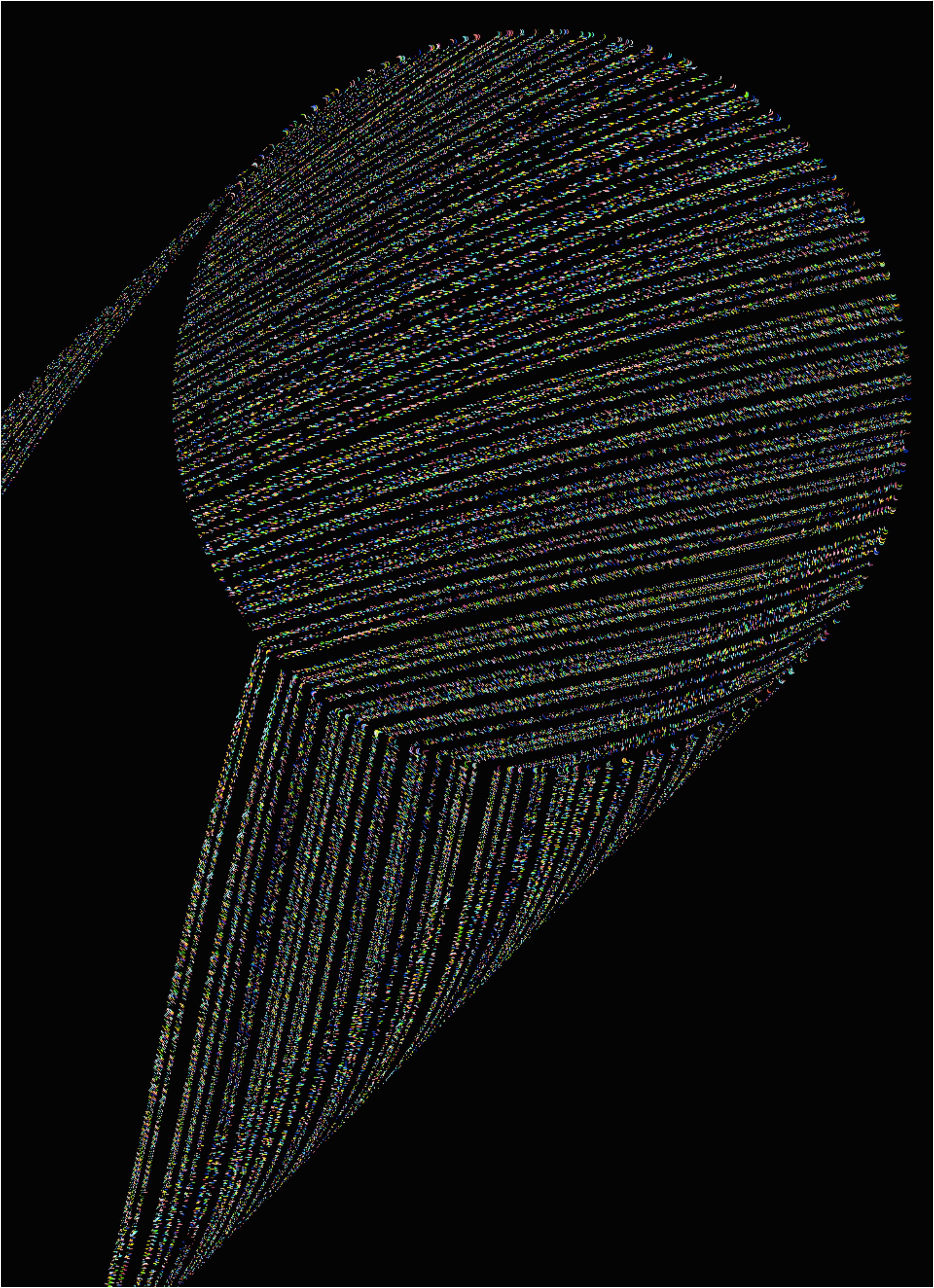}}
\caption{A beam composed of six 12$p_{Es}$ particles. Simulation is shown in a vertical position to get a better view of particles' trajectories. Interval between first and last particles can be any number {\sf mod} 14.}
\label{CTSrightArea}
\end{figure}

\begin{description}
\item[{\bf Left part}] \hfill \\ 
Periodic area handle beams of three trains of four $p_{A^4}$ particles, travelling from the left side with a constant velocity of $2/3$. This ring has 30,640 cells, the minimum interval between trains of particles is  649 copies of ether. Each beam of $p_{A^4}$ can have  three possible phases. The sequence of phases is periodic and fixed sequentially: $\{649e$-$4A^4$(F$_{i})\}^*$, for $1 \leq i \leq 3$ (Fig.~\ref{colliderDiagram} left area). Fig.~\ref{CTSleftArea} shows a simulation of these periodic beams of $4p_{A^4(F_{i})}$ particles.

\begin{figure}[th]
\centerline{\includegraphics[scale=.364]{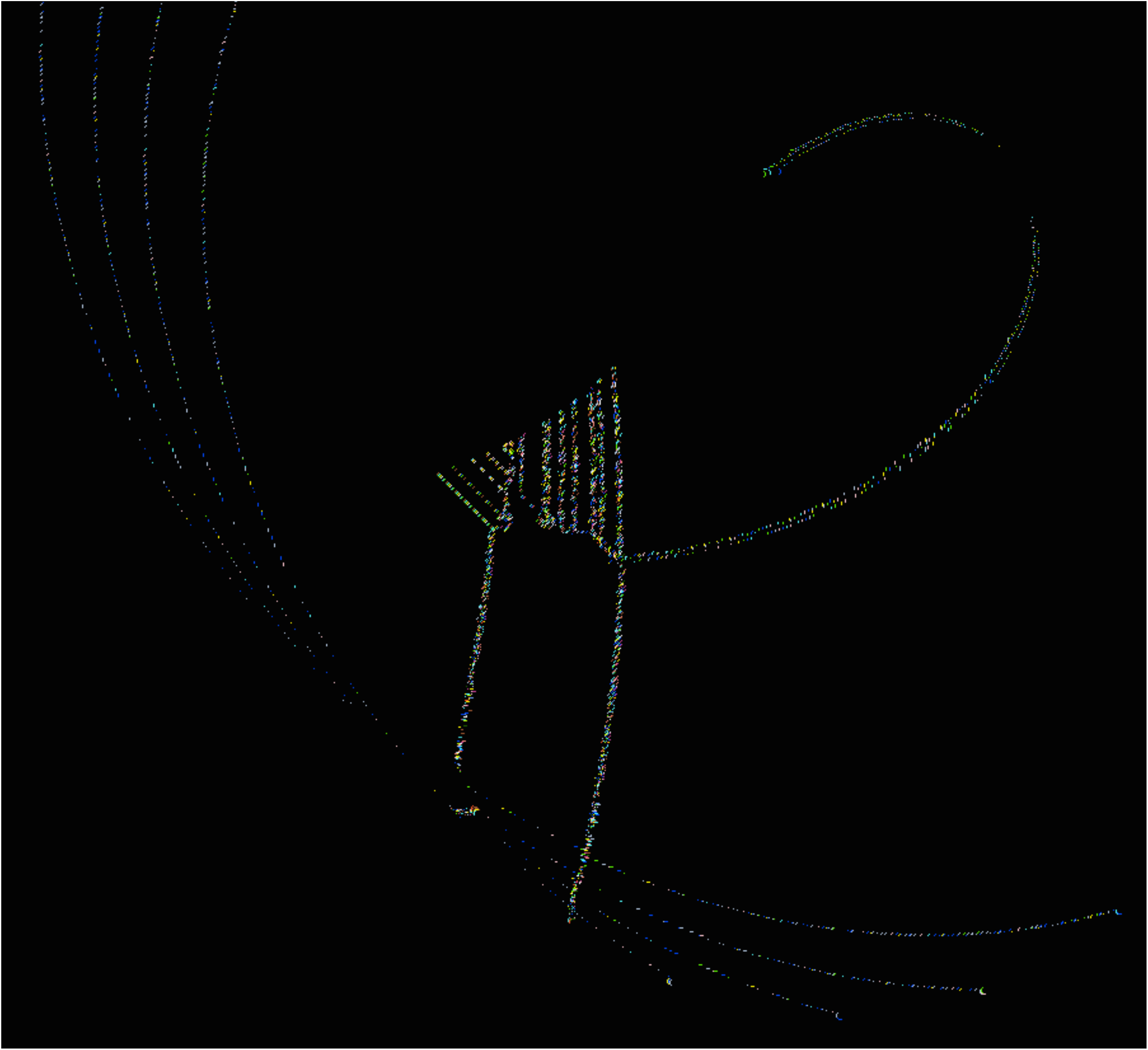}}
\caption{First stage of collisions of the cyclic tag system. Solitonic interactions take place  4$p_{A^4(F_3)}$ and two $p_{\bar{E}}$ particles. First symbol `1' on the type is deleted (center). The the first separator is read and deleted. }
\label{CTS1}
\end{figure}

\item[{\bf Right part}] \hfill \\ 
Periodic area handle beams of six trains of 12$E$'s particles ($p_{E^n}, p_{\bar{E}}$), travelling from the right side with a constant velocity of $-4/5$. There are 12 particles related to a perfect square with $13,500^2$ possibilities to arrange inputs into the main collider. Interval between 12 particles is {\sf mod} 14. Figure~\ref{CTSrightArea} shows the whole set of 72 $p_{Es}$ particles. The set contains leaders and separator components, and beams of particles that introduce `0's and `1's on  the tape. 


\begin{figure}[th]
\centerline{\includegraphics[scale=.39]{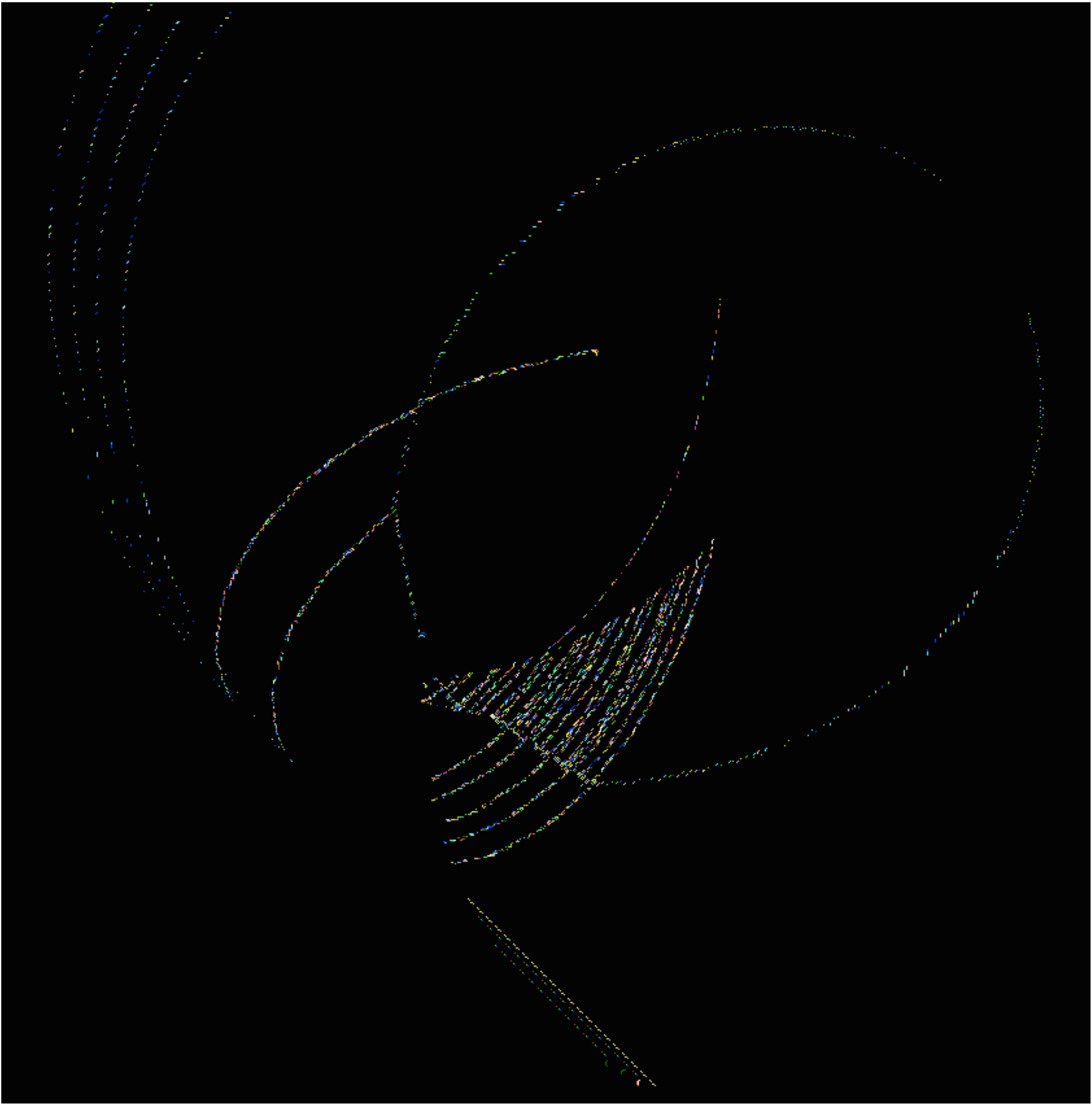}}
\caption{This snapshot shows when a `1' is introduced in the type. 
A second beam of 12 $p_{\bar{E}}$ particles is coming to leave just spaced four $p_{\bar{E}}$ particles, these particles collide with one of 4$p_{A^4(F_3)}$ particles. The result is four 4$p_{Cs}$ particles at the bottom of the simulation that represent one `1' in the cyclic tag system type.}
\label{CTS2}
\end{figure}

\item[{\bf Center}] \hfill \\
Initial state of particles starts with a `1' on the type of the cyclic tag system. Fig.~\ref{CTS1} shows the first stage of the collider. The system start with one `1' in the type (four vertical $p_{Cs}$ particles), they are static particles that wait for the first beam of $p_{Es}$ particles to arrive at the right side to delete this input and decode the next inputs. In this process two solitons emerge, but they do not affect the system and the first beam of $4p_{A^4(F_{3})}$ particles without changing their states.

\begin{figure}[th]
\centerline{\includegraphics[scale=.406]{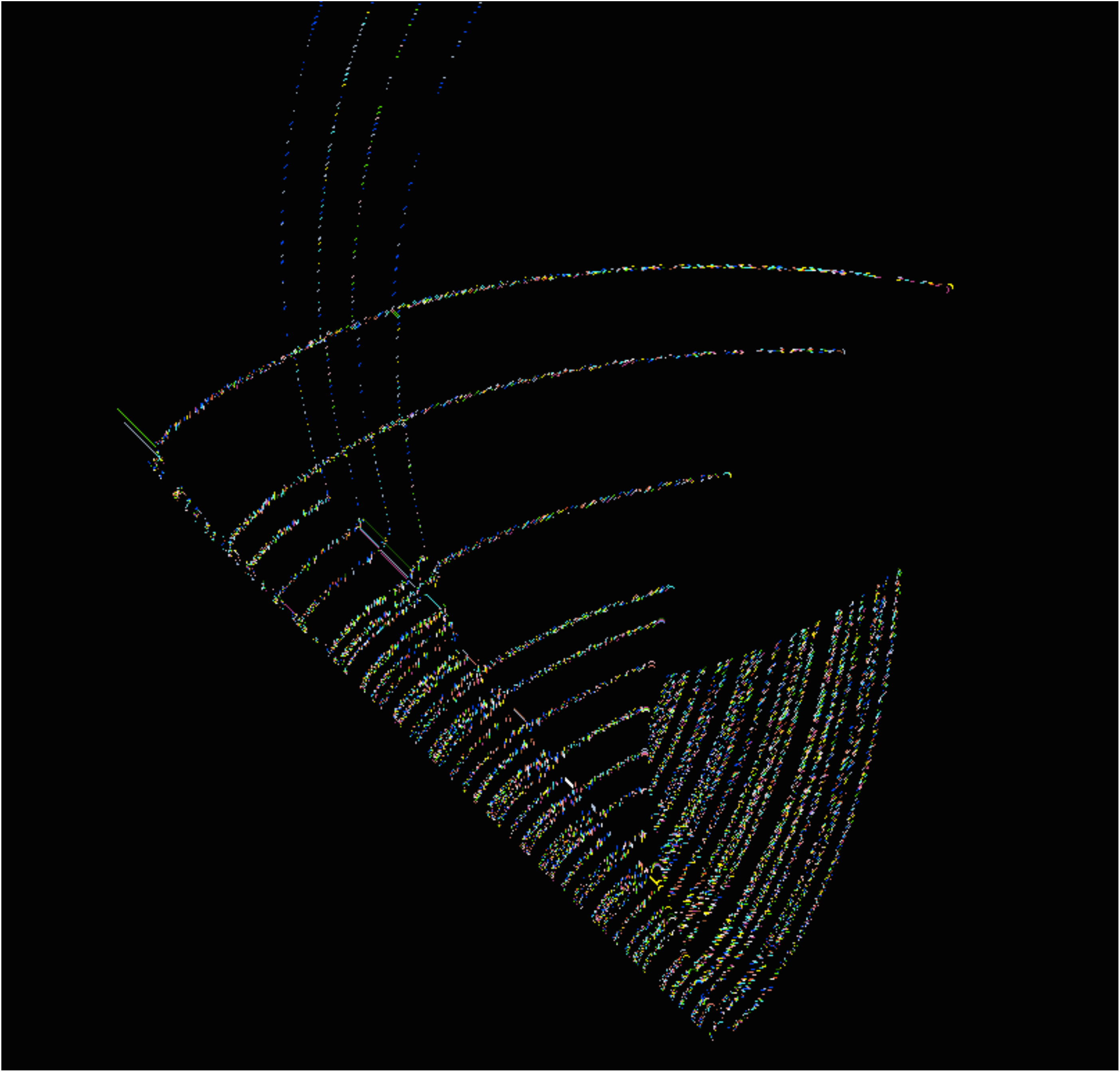}}
\caption{This snapshot shows how a sequence of values `0' and `1' is precoded. You can see sequences of `0's and `1's, and $p_{Es}$ particles travelling to from the left to the right.}
\label{CTS3}
\end{figure}

Figure~\ref{CTS2} shows how a second symbol `1' is introduced in the collider. A leader component is deleted and the second binary data is prepared to collide later with the first beam of 4$p_{A^4(F_3)}$ particles. Finally, the second `1' is represented for the vertical particles, as shown at the bottom of Fig.~\ref{CTS2}.

\begin{figure}[th]
\centerline{\includegraphics[scale=.405]{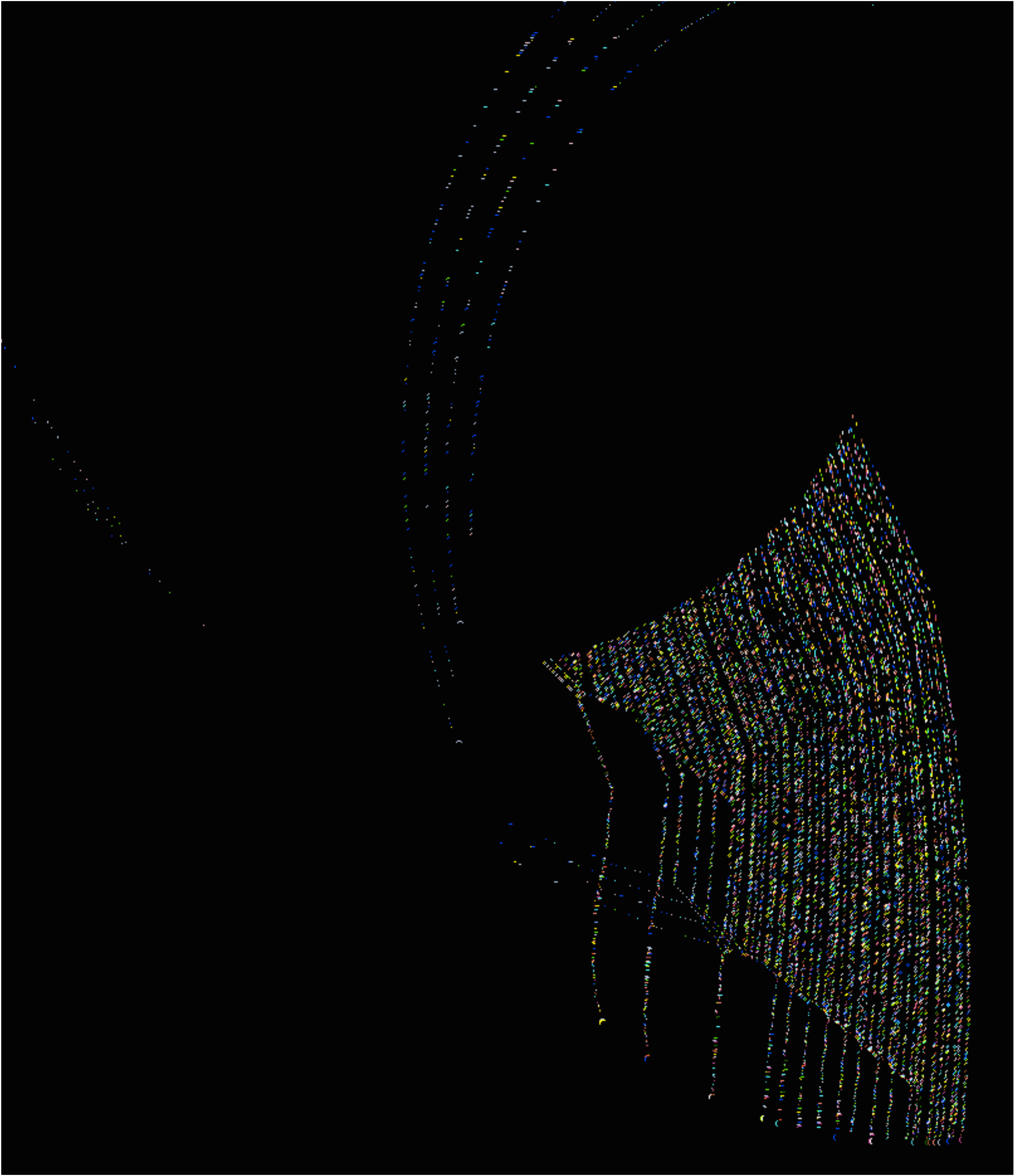}}
\caption{This snapshot shows from another angle how binary values are introduced in the cyclic tag system. We can also see how a number of values are prepared to collide with beams of 4$p_{A^4(F_i)}$ particles at the end of simulation.}
\label{CTS4}
\end{figure}

Figure~\ref{CTS3} shows how further symbols `0' and `1' are introduced in the system. They are coded with $p_{\bar{Es}}$ particles. Before the current `1' is introduced with 4$p_{A^4(F_3)}$ particles, the next set of 4$p_{A^4(F_3)}$ particles is prepared in advance.

\begin{figure}[th]
\centerline{\includegraphics[scale=.405]{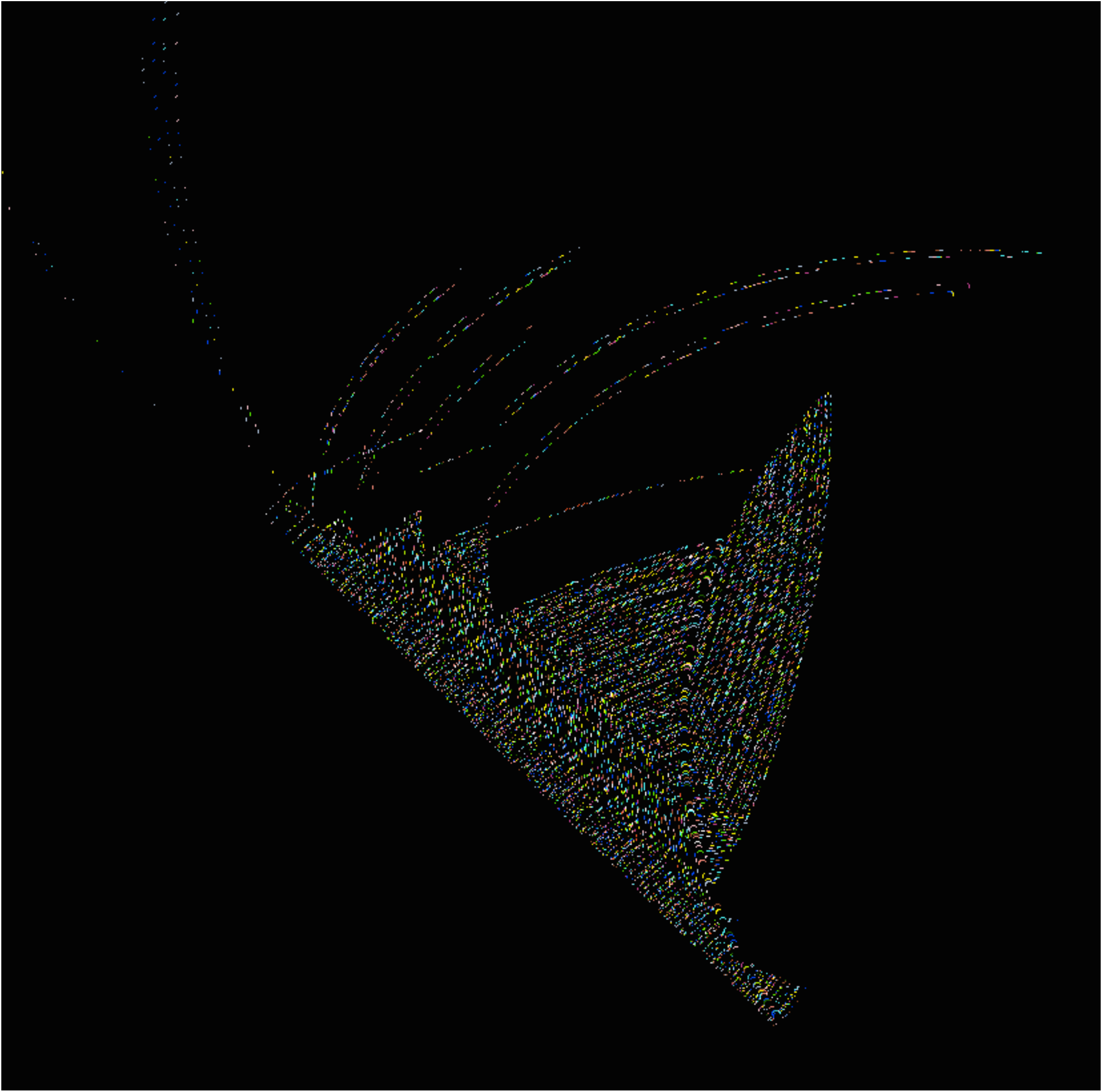}}
\caption{This evolution displays a full cycle of beams of $p_A$ and $p_{Es}$ particles. In this snapshot we can see all necessary operations in the cyclic tag system: input values, deleting block of values, particles like solitons, and the next stage of the collider.}
\label{CTS7}
\end{figure}

Figure~\ref{CTS4} shows the largest stage of the collider's working. A second beam of 4$p_{A^4(F_1)}$ arrives. More beams of $p_{Es}$ particles are introduced. Figure~\ref{CTS7} displays a full cycle of beams of $p_A$ and $p_{Es}$ particles. All operations are performed at least once. The next set of particles is ready to continue with the next stage of the computation.
\end{description}

\section{Discussion}

The CA collider is a viable prototype of a collision-based computing device. It well compliments existing models of computing circuits  based on particle collisions~\cite{heybook, finitecomp, fredkinToff, casupercomp, mills_2008, bumps, compuniv, supercacomp}. How complex is our design? With regarding to time complexity, rule simulates Turing machine a polynomial time and any step of rule 110 can be predicted in a polynomial time~\cite{compler110}. As to space complexity, left cyclotron in the collider is made of 30,640 cells and the right cyclotron of 5,864 cells. The main collider should have 61,280 cells to implement a full set of reactions; however, it is possible to reduce the number of cells in the main collider, because the first train of 4$p_{A^4(F_i)}$ particles needs just 10,218 cells; and subsequent trains can be prepared while initial data are processed. Thus, the simulated collider have just thousands of cells not millions.The space complexity of the implemented cyclic tag systems has been reduced substantially~\cite{ecauniversality, ankos, concreteview}. 

What are chances of implementing the CA collider model in physical substrates? A particle, or gliders, is a key component of the collider. The glider is a finite-state machine implementation of a propagation localisation. A solitary wave, or an impulse, propagating in a polymer chain could be a phenomenologically suitable analog of the glider. A wide range of polymer chains, both inorganic and organic, support solitons~\cite{davydov1977,heeger1988, scott1982, bredas1985, cnmac, cbc, unconv, noncc, massiveparallel, molsol, compsol}. We believe actin filaments could make the most suitable  substrate for implementation of a cyclic tag system via linked rings of CA colliders.
 
An actin filament is a double spiral helix of globular protein units. Not only actin is a key element of a cell skeleton, and is responsible for a cell's motility, but actin networks is a sensorial, information processing and decision making system of cells. In \cite{adamatzkymayne2015} we proposed a model of actin filaments as two chains of one-dimensional binary-state semi-totalistic automaton arrays. We show that a rich family of travelling localisations is observed in automaton model of actin, and many  of the localisation observed behave similarly to gliders in CA rule 110. The finite state machine model has been further extended to a quantum cellular automata model in \cite{siccardiquatum2015}. We have shown that quantum actin automata can perform basic operations of Boolean logic, and implemented a binary adder. To bring more `physical' meaning in our actin-computing concept we also employed the electrical properties of imitated actin filaments --- resistance, capacitance, inductance --- and found that it is possible to implement logical gates via interacting voltage impulses~\cite{siccardiRLC}; voltage impulses in non-linear transmission wires are analogs of gliders in 1D CA. Clearly, having just actin is not enough: we must couple rings together, arrange physical initiation of solitons and their detection, and solve myriad of other  experimental laboratory problems. That will be a scope of further studies.

\section{Additional stuff: video simulations}

\begin{itemize}
\item {\bf Cyclic left-side of particles in a cyclic tag system in ECA rule 110}. \\ URL: \url{https://youtu.be/HFJlbz7qATg}
\item {\bf Cyclic right-side of particles in a cyclic tag system in ECA rule 110}. \\ URL: \url{https://youtu.be/kciEa6cF1QQ}
\item {\bf A computation in a cellular automaton collider rule 110}. \\ URL: \url{https://youtu.be/i5af0tQiVd4}
\item {\bf Three glider guns evolving in a virtual collider in ECA rule 110}. \\ URL: \url{https://youtu.be/JElxQt32Odc}
\end{itemize}

\end{document}